\documentclass{article}

\usepackage{arxiv}
\usepackage{multirow}

\usepackage[utf8]{inputenc} % allow utf-8 input
\usepackage[T1]{fontenc}    % use 8-bit T1 fonts
\usepackage{hyperref}       % hyperlinks
\usepackage{url}            % simple URL typesetting
\usepackage{booktabs}       % professional-quality tables
\usepackage{amsfonts}       % blackboard math symbols
\usepackage{nicefrac}       % compact symbols for 1/2, etc.
\usepackage{microtype}      % microtypography
\usepackage{lipsum}

\usepackage{mathptmx}
\usepackage{threeparttable}
\usepackage{graphicx}

\title{Extracting candidate factors affecting long-term trends of student abilities across subjects\\\thanks{This work was supported by JSPS KAKENHI Grant Number 20H00098.}}

\author{
  Satoshi Takahashi\\
  College of Science and Engineering\\
  Kanto Gakuin University\\
  0000-0002-1067-6704\\
  \texttt{satotaka@kanto-gakuin.ac.jp} \\
  \And
  Hiroki Kuno\\
  Department of Computational Intelligence and Systems\\
  Tokyo Institute of Technology\\
  \And
  Atsushi Yoshikawa\\
  Department of Computational Intelligence and Systems\\
  Tokyo Institute of Technology\\
  0000-0001-7020-5085\\
  \texttt{at\_sushi\_bar@dis.titech.ac.jp}\\
}

\begin{document}
\maketitle

\begin{abstract}
  Long-term student achievement data provide useful information to formulate the research question of what types of student skills would impact future trends across subjects. However, few studies have focused on long-term data. This is because the criteria of examinations vary depending on their designers; additionally, it is difficult for the same designer to maintain the coherence of the criteria of examinations beyond grades. To solve this inconsistency issue, we propose a novel approach to extract candidate factors affecting long-term trends across subjects from long-term data. Our approach is composed of three steps: Data screening, time series clustering, and causal inference. The first step extracts coherence data from long-term data. The second step groups the long-term data by shape and value. The third step extracts factors affecting the long-term trends and validates the extracted variation factors using two or more different data sets. We then conducted evaluation experiments with student achievement data from five public elementary schools and four public junior high schools in Japan. The results demonstrate that our approach extracts coherence data, clusters long-term data into interpretable groups, and extracts candidate factors affecting academic ability across subjects. Subsequently, our approach formulates a hypothesis and turns archived achievement data into useful information.
\end{abstract}

% keywords can be removed

\section{Introduction}
Educational interventions have a long-term effect, and attempts have been made to use early-stage intervention to improve the learning ability of lower and upper secondary school students \cite{Hwang2018}. Additionally, certain abilities require long-term treatment to generate improvement \cite{Vaughn2008}. Educational interventions also have an impact across subjects \cite{Mullis2011}. The school curriculum is designed to assume that students comprehend their previous grade's curriculum, and a lack of comprehension at an early stage would harm students' abilities in other subjects in the future.

Few studies, however, have focused on long-term mutual interaction across subjects. Analysis of this long-term mutual interaction could identify students who require help early and the topics that teachers should pay attention to across subjects.

However, utilizing long-term student achievement data comes with difficulties. Schools conduct many examinations, including school achievement tests and public educational assessments. These have different purposes, and their criteria and assumptions vary depending on their designers. Additionally, the designers have to create examinations corresponding to students' grades; hence, it is difficult for the same designer to maintain the coherence of the criteria of examinations between different grades. This inconsistency makes it difficult to apply previous time-series analyses to the long-term data without ingenuity. Further, this difficulty is problematic for schoolteachers and educational policymakers attempting to extract candidate factors affecting long-term trends in student ability across subjects for educational policymaking.

We propose a novel approach for extracting the candidate factors affecting long-term trends in students' ability across subjects to solve this issue. First, we discuss the previous research on long-term educational effects and mutual interaction across subjects. Then, we propose a novel approach composed of three steps: data screening, time series clustering, and causal inference. Finally, we conduct experiments evaluating the proposed approach.

\section{Related work}

\subsection{Long-term educational effects}
Many studies have focused on the long-term educational effects of interventions. For example, Merki and Oerke \cite{Merki2017} focused on the long-term effects of implementing state-wide exit exams over five years using a multilevel regression analysis. They then reveal that state-wide exams have positive effects on teaching practices and students' motivation. Droop, van Elsäcker, Voeten, and Verhoeven \cite{Droop2016} examined the effects of sustained strategic reading instruction of third and fourth graders and found positive effects on knowledge of reading strategies and reading comprehension. Watts, Clements, Sarama, Wolfe, Spitler, and Bailey \cite{Watts2017} focused on why early educational intervention effects typically fade in the years following treatment. They find that educational intervention can affect the underlying processes in children's mathematical development by inducing the transfer of knowledge.

The educational effect has a time lag, and some interventions' effects become apparent after a specific time. For this reason, a student's achievements and behaviors should be observed for an extended period. Rousseau \cite{Rousseau2006}, for instance, has suggested that the benefits of social promotions will increase the employment success rate and prevent drug use a few years later. Cawley, Heckman, and Vylacil \cite{Cawley1998} analyzed the contribution to the economy of the return to education. They demonstrated that the college-high school premium increased in the mid-80s for high-ability young people. Cunha and Heckman \cite{Cunha2009} analyzed the relationship between ``cognitive and non-cognitive capability'' and investment. They demonstrated that investment can enhance a child's capability and self-productivity and that it is relatively more productive at some stages of a child's life cycle; for example, investment is relatively more effective in increasing adult earnings when made in children aged 8--9 to 10--11 compared to children aged 6--7 to 8–9 and 10–11 to 12–13.

Time-series analyses can be a powerful tool for examining the long-term effects of education practice \cite{Aghabozorgi2015,Liao2005}. Kaufman and Segura-Ubiergo \cite{Kaufman2001} analyzed the relationship between social security transfers and ``health and education expenditures'' with a time-series cross-sectional analysis. They used data for Latin American countries such as central government spending, GDP, and public expenditures on health care, education, and social security programs from 1973 to 1997. Loening \cite{Loening2002} investigated the impact of human capital on economic growth in Guatemala. He applied a time series analysis to error-correction methodology, using data such as public spending on education in 1995, average years of schooling in 1996, and primary school net enrollment in 1997. Based on this analysis, he claims that a better-educated labor force appears to have a significant positive impact on economic growth via factor accumulation.

\subsection{Mutual interaction across subjects}
The importance of relationships across subjects has been noted, with a call for the reform of the school curriculum to reflect it \cite{Boss2008}. In particular, various studies have demonstrated that writing skills have an impact on mathematics \cite{Boss2008,Borasi1990}. Shaftel, Belton-Kocher, Glasnapp, and Poggio \cite{Shaftel2006} examined the relationship between mathematics and linguistic characteristics and illustrated that the difficulty level of mathematics vocabulary affects performance. Freitag \cite{Freitag1997} argued that reading skills in mathematics are necessary for students to comprehend problems written as text but can cause problems with their comprehension of how to solve the problem and represent their ideas in writing. Notably, mathematics includes symbols and formulas written as text. Hence, students with different primary language skills often face difficulties in mathematics lessons \cite{Bohlmann2002}.

The International Association for the Evaluation of Educational Achievement (IEA) has conducted an international survey of students' mathematics and science skills via the Trends in International Mathematics and Science Study (TIMSS) since 1995. Mullis, Martin, and Foy \cite{Mullis2011} developed an indicator to assess the necessary reading skill level in TIMSS items based on the number of words, vocabulary, symbolic language, and visual display. With this indicator, they analyzed the results of fourth-grade students in the 2011 TIMSS, in which over 600,000 students from 34 countries participated. They concluded that in most countries, students who have a high reading ability have a good record of items requiring high reading skills.\\
Several studies have also conducted surveys on the relationships between subjects, such as science and reading skills \cite{DiGisi1992}, and music and reading skills \cite{Anvari2002,Hansen2002,Zinar1976}.

\subsection{Linking individual tests scores}
Schools conduct many assessments, including the National Assessment of Educational Progress (NEAP), the Programme for International Student Assessment (PISA), and TIMSS, as well as their own exams. The NEAP, PISA, and TIMSS are public assessments designed from individual educational policy criteria, while teachers create school tests to assess students' comprehension of content from the previous year. The public assessments and the schools' tests thus have different, individual purposes. Brown, Micklewright, Schnepf, and Waldmann \cite{Brown2007} compared several countries' scores from the PISA, TIMSS, International Adult Literacy Survey, and Progress in International Reading Literacy Study. They demonstrated that the correlations within the survey for different subjects are higher than those between surveys for similar subjects and concluded, therefore, that it is worth considering the nature of the distributed surveys themselves.

Many studies have attempted to solve this issue, for example, by trying to connect different tests. Kolen and Brennan \cite{Kolen2014} demonstrated a connection between the American College Testing, Science Reasoning test, and the Iowa Tests of Educational Development Analysis of Science Materials test. Liu and Walker \cite{Liu2007} connected the NAEP, International Assessment of Educational Progress, the Armed Services Vocational Aptitude Battery, and the North Carolina End-of-Grade Tests. These approaches are referred to as ``linking,'' and they focus on translating one test score into other tests' score systems \cite{Kolen2014}.

\subsection{Summary}
Many studies, as well as real-life practices, have illustrated that early-stage interventions have a positive impact on long-term development and claim that relationships across subjects are important. However, only a few studies have focused on the long-term trends of students' achievements, such as Stanley, Petscher, and Catts \cite{Stanley2018}, who examined the relationship between reading skills in kindergarten and those in tenth grade. Bodovski and Youn \cite{Bodovski2011} examined the relationship between first-grade students' behaviors and their reading and mathematics achievements in fifth grade, while Sparks, Patton, and Murdoch \cite{Sparks2014} examined the relationship of reading skills over ten years from first to tenth grade.

The lack of research is largely due to the significant difficulty in analyzing long-term student performance. Schools have conducted many assessments; however, public assessments and the schools' tests have distinct purposes. Additionally, it is difficult for the same designer to consider test factors beyond grades continually. As a result, tests have variations and inconsistencies among them. Ding \cite{Ding2009} also pointed out that we cannot extract good conclusions from evaluation data with mismatched analysis purposes. In addition, previous research has focused on translating scores from one test into scores on other tests. Although we focus on long-term student performance, our purpose is not to translate a given test's scores into future or past tests' scores, and we cannot mix and examine different types of tests without the ingenuity of analyzing long-term student performance.\\

\clearpage
\section{Proposed approach}
We propose a novel approach to extract candidate factors affecting long-term trends of students' abilities across subjects. The long-term data that our approach targeted had the following features: (1) They were measured by different exams in each period; (2) they comprised individual students' achievement data to find individual students' trends; (3) they included the subject score to find moving patterns; (4) they included each item's scores to extract variation factors; (5) they included test data sets across subjects to extract the long-term mutual interaction; and (6) they included different data sets to validate the extracted variation factors.

There were some issues in applying student achievement data for a time series analysis. The first issue was that long-term achievement data sometimes lack coherence. To solve this issue, our approach extracted coherence data from long-term data. Specifically, our approach utilized individual students' ranks in each test and assessed the coherence of individual students' ranks among the tests. Then, our approach adopted the test data sets for which coherence was stable because it was difficult to believe that a large number of students' scores would suddenly change at the same time; instead, it was more likely that the evaluation criteria of a given test changed and the long-term data lost coherence.

The second issue was that there would be many trend patterns in the long-term data. For example, some students gradually increase their scores; some students gradually lower their scores; and some students have lower scores at first but later raise their scores. These patterns would have different variation factors, and they cannot be handled together. Our approach utilized time series clustering to group students by trend patterns. It should be noted that the grouping method has to consider both the timeline shape and value because even trends of the same shape have different factors (e.g., one might raise its score from the bottom line and another from the middle line).

The third issue was that our approach had to extract factors affecting the long-term trends. In general, the causal inference method needs more than two different groups. Therefore, our approach applied the causal inference method to groups such that their scores were the same initially but different later. Further, the validation of the time series clustering and causal inference is essential. Hence, our approach used two or more different data sets individually and was validated by extracting the same results from those data sets.

\begin{figure}[htbp]
  \centering
  \includegraphics[width=10cm]{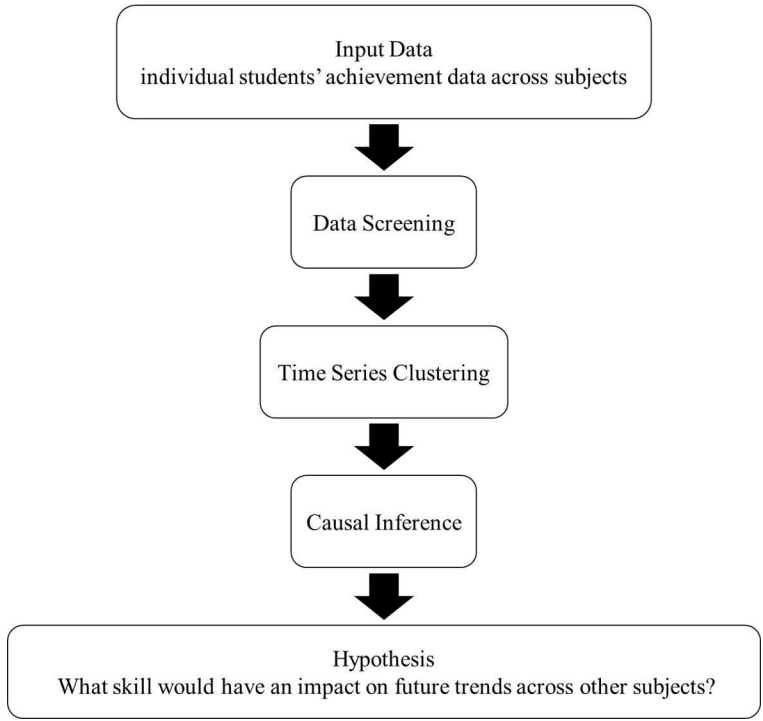}
  \caption{The procedures of our approach}
  \label{fig:fig1}
\end{figure}

Figure \ref{fig:fig1} illustrates our approach. The first step corresponded to the first issue: Data screening. In this step, our approach extracted coherence data from the long-term data. The second step corresponded to the second issue: Time series clustering. In this step, our approach grouped the long-term data by score changing patterns. The third step corresponded to the third issue: Causal inference. In this step, our approach extracted candidate factors affecting the long-term trends and validated extracted variation factor uses with two or more different data sets.

\subsection{Data screening step}
In this step, our approach extracted the coherence data series and then adopted a correlation analysis. Correlation analysis is a statistical method to evaluate the strength of a relationship between two variables and is used for an interval, a ratio, and an ordinal scale. In many cases, Pearson's correlation coefficient is used for an interval and a ratio scale, and Spearman's rank correlation coefficient is used for an ordinal scale.

\begin{figure}[htbp]
  \centering
  \includegraphics[width=10cm]{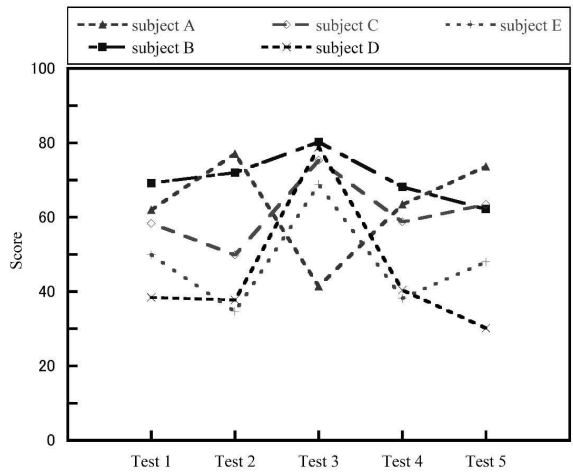}
  \caption{Example of an individual test score}
  \label{fig:fig2}
\end{figure}

With a correlation analysis, our approach extracted the test data sets for which coherence was stable. Figure \ref{fig:fig2} illustrates an example of time series student achievement data in five subjects (subject A, subject B, subject C, subject D, and subject E). The example data include scores of tests 1, 2, 3, and 4, and the chronological order is test 1, test 2, test 3, and test 4.

Table \ref{table:table1} illustrates the correlation analysis among the tests. Then, our approach focused on the correlation coefficients ($r$) between tests of two consecutive times: between test 1 and test 2, $r$ was 0.84; between test 2 and test 3, $r$ was 0.50; between test 3 and test 4, $r$ was 0.23; and between test 4 and test 5, $r$ was 0.83.

The correlation coefficients of test 3 were very low compared to the other correlation coefficients. Additionally, between test 2 and test 4, $r$ was 0.92 and was higher than $r$ between test 2 and test 3 and between test 3 and test 4. When the evaluation criteria of the test changed, the Pearson's correlation coefficient became low. Therefore, our approach judged that the evaluation criteria of test 3 were different from those of the other tests, and as a result, our approach excluded test 3. The results of the correlation analysis among tests without test 3 were as follows: between test 1 and test 2, $r$ was 0.84; between test 2 and test 4, $r$ was 0.92; and between test 4 and test 5, $r$ was 0.83. Our approach thus moved to the next step without test 3.

\begin{table}[htb]
  \begin{center}
    \caption{Example of correlation analysis among tests}
    \begin{tabular}{cccccc} \hline
             & Test 1 & Test 2 & Test 3 & Test 4 & Test 5 \\ \hline
      Test 1 &        & 0.84   & 0.23   & 0.91   & 0.88   \\ \hline
      Test 2 &        &        & 0.50   & 0.92   & 0.81   \\ \hline
      Test 3 &        &        &        & 0.23   & 0.58   \\ \hline
      Test 4 &        &        &        &        & 0.83   \\ \hline
      Test 5 &        &        &        &        &        \\ \hline
    \end{tabular}
    \label{table:table1}
  \end{center}
\end{table}

\subsection{Time Series Clustering step}
In this step, our approach utilized time series clustering to group students by score changing patterns. Cluster analysis is a method to divide a data set into groups based on similarity. It is roughly classified into hierarchical clustering and non-hierarchical clustering. Hierarchical clustering divides a data set stepwise based on similarity and demonstrates the hierarchical relationship between the groups using a dendrogram. Non-hierarchical clustering divides data into a specific number of groups defined in advance. Our approach could have adopted both types of cluster analysis.

Our approach had to consider both the timeline shape and value of the long-term data. The combination of Dynamic Time Warping (DTW) and k-means clustering was one of the methods appropriate for such a purpose \cite{Sakoe1978}. Many other suitable methods had been proposed, such as Euclidean, Longest Common Sub-Sequence \cite{Vlachos2002} and Minimal Variance Matching \cite{Latecki2005}.

In general, it is not easy to evaluate the results of clustering in the absence of data labels \cite{Liao2005}. The appropriate clusters depend on the user and domain, and this is subjective \cite{Aghabozorgi2015}. Thus, our approach needed to choose the suitable number of the group, considering that the clusters were interpretable from educational criteria.

Our approach used two or more different data sets individually and validated that the same results were extracted from those data sets. Table \ref{table:table2} illustrates a successful example of time series clustering where both data set 1 and data set 2 were clustered into the same cluster types: type A and type B. By contrast, Table \ref{table:table3} illustrates an example of a failure, where data set 1 and data set 2 were clustered into different cluster types.

\begin{table}[htbp]
  \centering
  \caption{Success example of time series clustering}
  \begin{tabular}{cc} \hline
    Datasets                  & Cluster type   \\ \hline
    \multirow{3}{*}{Dataset1} & Cluster type A \\ \cline{2-2}
                              & Cluster type B \\ \cline{2-2}
                              & Cluster type C \\ \hline
    \multirow{3}{*}{Dataset2} & Cluster type A \\ \cline{2-2}
                              & Cluster type B \\ \cline{2-2}
                              & Cluster type C \\ \hline
  \end{tabular}
  \label{table:table2}
\end{table}

\begin{table}[htbp]
  \centering
  \caption{Failure example of time series clustering}
  \begin{tabular}{cc} \hline
    Datasets                  & Cluster type   \\ \hline
    \multirow{3}{*}{Dataset1} & Cluster type A \\ \cline{2-2}
                              & Cluster type B \\ \cline{2-2}
                              & Cluster type C \\ \hline
    \multirow{3}{*}{Dataset2} & Cluster type D \\ \cline{2-2}
                              & Cluster type E \\ \cline{2-2}
                              & Cluster type F \\ \hline
  \end{tabular}
  \label{table:table3}
\end{table}

\clearpage
\subsection{Causal Inference step}
In this step, our approach extracted candidate factors affecting long-term trends across subjects. Causal inference is the method for identifying the causes of a phenomenon. Popular causal inference methods include multivariate logistic regressions, structural equation modeling, and regression analyses \cite{Han2011}. For this paper, we used a multivariate logistic regression \cite{Glonek1995}.

Our approach used two or more different data sets individually and validated that the causal inference method extracted the same causal inference from those data sets. Table \ref{table:table4} illustrates a successful example of causal inference, where data set 1 and data set 2 have the same variation factors: factor A and factor B. In contrast, Table \ref{table:table5} illustrates a failed example of causal inference, where data set 1 and data set 2 did not have the same variation factors.

\begin{table}[htbp]
  \centering
  \caption{Success example of causal inference}
  \begin{tabular}{cc} \hline
    Datasets                                          & Results of Causal Inference \\ \hline
    \multirow{3}{*}{Cluster type A and B of Dataset1} & Variation factor A          \\ \cline{2-2}
                                                      & Variation factor B          \\ \cline{2-2}
                                                      & Variation factor C          \\ \hline
    \multirow{3}{*}{Cluster type A and B of Dataset2} & Variation factor A          \\ \cline{2-2}
                                                      & Variation factor B          \\ \cline{2-2}
                                                      & Variation factor D          \\ \hline
  \end{tabular}
  \label{table:table4}
\end{table}

\begin{table}[htbp]
  \centering
  \caption{Failure example of causal inference}
  \begin{tabular}{cc} \hline
    Datasets                                          & Results of Causal Inference \\ \hline
    \multirow{3}{*}{Cluster type A and B of Dataset1} & Variation factor A          \\ \cline{2-2}
                                                      & Variation factor B          \\ \cline{2-2}
                                                      & Variation factor C          \\ \hline
    \multirow{3}{*}{Cluster type A and B of Dataset2} & Variation factor D          \\ \cline{2-2}
                                                      & Variation factor E          \\ \cline{2-2}
                                                      & Variation factor F          \\ \hline
  \end{tabular}
  \label{table:table5}
\end{table}

\clearpage
\section{Evaluation Experiment 1}
We conducted an experiment to evaluate our approach. The input data were the student achievement data from five public elementary schools and four public junior high schools and included individual students' time-series data from fourth to ninth grade between 2014 and 2018. Japanese elementary schools are from first to sixth grade, and Japanese junior high schools are from seventh to ninth grade.

The input data also included national language and mathematics achievement tests from two different organizations. The tests for grades four, five, seven, and eight were organization A's achievement tests. The tests for grades six and nine were organization B's achievement tests.

We divided the input data into two groups (Tables \ref{table:table6} and \ref{table:table7}). The students of group 1 were in the fourth grade in 2014, and the students of group 2 were in the fifth grade in 2014. The schools of both groups were the same, and all students took all achievement tests. The number of group 1 students was 168, and the number of group 2 students was 201.

The achievement tests for organization A were conducted during grades four, five, seven, and eight. Test subjects were national language and mathematics, and the tests covered content that students learned in the previous year (e.g., the test for fourth grade covered content learned in third grade).

The achievement test data included individual test items, their topics, individual students' points, individual students' answers, and individual students' deviation scores, which were represented by formula (1):

\begin{equation}
  T_i=\frac{10(x_i - \mu)}{\sigma}+50.
\end{equation}

Where $T$ is the individual deviation score, $x$ is the individual achievement test score, $i$ is the student $i$, $\mu$ is the arithmetic mean of the achievement test, and $\sigma$ is the standard deviation of the achievement test.

The achievement tests for organization B were conducted in grades six and nine. The test subjects were two types of national language and mathematics tests: national language types A and B and mathematics types A and B. The tests covered content that students learned in the previous year. The achievement test data included individual test items, their topics, and individual students' answers, but not individual students' points and deviation scores.

\begin{table}[htb]
  \begin{center}
    \caption{Student achievement data on national language exams}
    \begin{tabular}{cccccc} \hline
      Group                      & 2014           & 2015                       & 2016           & 2017           & 2018                       \\ \hline
      Group 1                    & [Org. A] 4 NL. & [Org. A] 5 NL.             &
      \begin{tabular}{c} [Org. B] 6 NL. A \\ {[Org. B] 6 NL. B} \end{tabular} &
      [Org. A] 7 NL.             & [Org. A] 8 NL.                                                                                             \\ \hline
      Group 2                    & [Org. A] 5 NL. & \begin{tabular}{c} [Org. B] 6 NL. A \\ {[Org. B] 6 NL. B} \end{tabular} & [Org. A] 7 NL. & [Org. A] 8 NL. & \begin{tabular}{c} [Org. B] 9 NL. A \\ {[Org. B] 9 NL. B} \end{tabular} \\ \hline
    \end{tabular}
    \\Note. Org. is an abbreviation for organization, NL. is an abbreviation for national language, NL. A. is an abbreviation for national language type A, and NL. B. is an abbreviation for national language type B.\\
    E.g., [Org. A] 4 NL. is an abbreviation for organization A's national language test for the fourth grade.
    \label{table:table6}
  \end{center}
\end{table}

\begin{table}[htb]
  \begin{center}
    \caption{Student achievement data of mathematics}
    \begin{tabular}{cccccc} \hline
      Group                      & 2014          & 2015                       & 2016          & 2017          & 2018                       \\ \hline
      Group 1                    & [Org. A] 4 M. & [Org. A] 5 M.              &
      \begin{tabular}{c} [Org. B] 6 M. A \\ {[Org. B] 6 M. B} \end{tabular} &
      [Org. A] 7 M.              & [Org. A] 8 M.                                                                                           \\ \hline
      Group 2                    & [Org. A] 5 M. & \begin{tabular}{c}[Org. B] 6 M. A\\ {[Org. B] 6 M. B} \end{tabular} & [Org. A] 7 M. & [Org. A] 8 M. & \begin{tabular}{c}[Org. B] 9 M. B\\ {[Org. A] 9 M. B} \end{tabular} \\ \hline
    \end{tabular}
    \\Note. Org. is an abbreviation for organization, M. is an abbreviation for mathematics, M. A is an abbreviation for mathematics type A, and M. B is an abbreviation for mathematics type B.

    \label{table:table7}
  \end{center}
\end{table}

\subsection{Data screening step}
We conducted a correlation analysis with the correct answer ratios in each achievement test. First, we analyzed group 1's results. Table \ref{table:table8} illustrates the results of the national language achievement tests, and Table \ref{table:table9} illustrates the results of the mathematics achievement tests. In the analysis of the results, Org. is an abbreviation for organization, NL. is an abbreviation for national language, NL. A. is an abbreviation for national language type A, and NL. B. is an abbreviation for national language type B. For example, [Org. A] 4 NL. is an abbreviation for organization A's national language test for the fourth grade.

In Table \ref{table:table8}, the rs among the consecutive tests for organization A were 0.76 or higher; [Org. A] 4 NL., [Org. A] 5 NL., [Org. A] 7 NL., and [Org. A] 8 NL. In contrast, the rs among the consecutive tests for organizations A and B were 0.62 or lower; [Org. A] 5 NL., [Org. B] 6 NL. A NL., [Org. B] 6 NL. B NL., and [Org. A] 7 NL. According to these results, the evaluation criteria between the tests for organizations A and B seemed different. Therefore, we excluded the achievement test for organization B.

In Table \ref{table:table9}, the rs among the consecutive tests for organization A were 0.72 or higher; [Org. A] 4 M., [Org. A] 5 M., [Org. A] 7 M., and [Org. A] 8 NL. In contrast, the rs among the consecutive tests for organizations A and B were 0.69 or lower; [Org. A] 5 M., [Org. B] 6 M. A, [Org. B] 6 M. B., and [Org. A] 7 M. According to these results, we considered the evaluation criteria between tests for organizations A and B to be different. We thus excluded the achievement test for organization B.

Finally, we extracted [Org. A] 4 NL., [Org. A] 5 NL., [Org. A] 7 NL., and [Org. A] 8 NL. from national language tests. The correlation coefficients were as follows: the $r$ between [Org. A] 4 NL. and [Org. A] 5 NL. was 0.81, the $r$ between [Org. A] 5 NL. and [Org. A] 7 NL. was 0.82, and the $r$ between [Org. A] 7 NL. and [Org. A] 8 NL. was 0.76.

Additionally, we extracted [Org. A] 4 M., [Org. A] 5, [Org. A] 7 M., and [Org. A] 8 M. from the mathematics tests. The correlation coefficients were as follows: the $r$ between [Org. A] 4 M. and [Org. A] 5 was 0.82, the $r$ between [Org. A] 5 M. and [Org. A] 7 M. was 0.85, and the $r$ between [Org. A] 7 M. and [Org. A] 8 M. was 0.72.

Then, we analyzed group 2's results. Table \ref{table:table10} illustrates the results of national language achievement tests; Table \ref{table:table11} illustrates the mathematics achievement test results. In Table \ref{table:table10}, the $r$ between [Org. A] 7 NL. and [Org. A] 8 NL. was 0.72, and the $r$ between [Org. A] 8 NL. and [Org. B] 9 NL. A was 0.70. By contrast, the other rs among consecutive tests were 0.68 or lower. Additionally, $r$ between [Org. A] 5 NL. and [Org. A] 7 NL. was 0.70. According to these results, we considered the evaluation criteria of [Org. B] 6 NL. A, [Org. B] 6 NL. B, [Org. A] 9 NL. A, and [Org. A] 9 NL. B to be different from the other tests. Thus, we excluded those tests; the extracted tests were [Org. A] 5 NL., [Org. A] 7 NL., and [Org. A] 8 NL.

In Table \ref{table:table11}, the rs among [Org. A] 7 M., [Org. A] 8, [Org. B] 9 M. A, and [Org. B] 9 M. B were 0.78 or higher. By contrast, the other rs among the consecutive tests were 0.69 or lower. Additionally, the $r$ between [Org. A] 5 M. and [Org. A] 7 M. was 0.79. According to these results, we considered the evaluation criteria among [Org. B] 6 M. A and [Org. B] 6 M. B to be different from the other tests. Thus, we excluded those tests, and the extracted tests were [Org. A] 5 M., [Org. A] 7 M., [Org. A] 8 M., [Org. B] 9 M. A, and [Org. B] 9 M. B.

Finally, we extracted [Org. A] 5 NL., [Org. A] 7 NL., [Org. A] 8 NL., and [Org. B] 9 NL. A from national language tests. The correlation coefficients were as follows: the $r$ between [Org. A] 5 NL. and [Org. A] 7 NL. was 0.70, the $r$ between [Org. A] 7 NL. and [Org. A] 8 NL. was 0.72, the $r$ between [Org. A] 8 NL. and [Org. B] 9 NL. A was 0.70.

Additionally, we extracted [Org. A] 5 M., [Org. A] 7, [Org. A] 8 M., [Org. B] 9 M. A, and [Org. B] 9 M. B from the mathematics tests. The correlation coefficients were as follows: the $r$ between [Org. A] 5 M. and [Org. A] 7 M. was 0.79, the $r$ between [Org. A] 7 M. and [Org. A] 8 M. was 0.78, the $r$ between [Org. A] 8 M. and [Org. B] 9 M. A was 0.85, and the $r$ between [Org. A] 8 M. and [Org. B] 9 M. B was 0.82.

The Time Series Clustering step required the same test type data sets to validate the results, so we selected the tests included in both groups 1 and 2. For example, when we focused on the national language achievement tests, our approach selected [Org. A] 5 NL., [Org. A] 7 NL., and [Org. A] 8 NL. Additionally, when we focused on the mathematics achievement tests, our approach selected [Org. A] 5 M., [Org. A] 7 M., and [Org. A] 8 M.

\begin{table}[htb]
  \begin{center}
    \caption{Group 1's results among achievement tests for national language}
    \begin{tabular}{ccccccc}
      \hline

                                 &
      \begin{tabular}{c}[Org. A]\\4 NL.\end{tabular} &
      \begin{tabular}{c}[Org. A]\\5 NL.\end{tabular} &
      \begin{tabular}{c}[Org. B]\\6 NL. A\end{tabular} &
      \begin{tabular}{c}[Org. B]\\6 NL. B\end{tabular} &
      \begin{tabular}{c}[Org. A]\\7 NL.\end{tabular} &
      \begin{tabular}{c}[Org. A]\\8 NL.\end{tabular}   \\\hline

      \begin{tabular}{c}[Org. A]\\4 NL.\end{tabular} &
                                 &
      $0.81^{\ast\ast}$          &
      $0.57^{\ast\ast}$          &
      $0.55^{\ast\ast}$          &
      $0.71^{\ast\ast}$          &
      $0.72^{\ast\ast}$            \\\hline

      \begin{tabular}{c}[Org. A]\\5 NL.\end{tabular} &
                                 &
                                 &
      $0.58^{\ast\ast}$          &
      $0.58^{\ast\ast}$          &
      $0.82^{\ast\ast}$          &
      $0.75^{\ast\ast}$            \\\hline

      \begin{tabular}{c}[Org. B]\\6 NL. A\end{tabular} &
                                 &
                                 &
                                 &
      $0.61^{\ast\ast}$          &
      $0.62^{\ast\ast}$          &
      $0.56^{\ast\ast}$            \\\hline

      \begin{tabular}{c}[Org. B]\\6 NL. B\end{tabular} &
                                 &
                                 &
                                 &
                                 &
      $0.60^{\ast\ast}$          &
      $0.58^{\ast\ast}$            \\\hline

      \begin{tabular}{c}[Org. A]\\7 NL.\end{tabular} &
                                 &
                                 &
                                 &
                                 &
                                 &
      $0.76^{\ast\ast}$            \\\hline

      \begin{tabular}{c}[Org. A]\\8 NL.\end{tabular} &
                                 &
                                 &
                                 &
                                 &
                                 &
      \\\hline
    \end{tabular}
    \\Note. $ ^{\ast\ast} p<.01$
    \label{table:table8}
  \end{center}
\end{table}

\begin{table}[htb]
  \begin{center}
    \caption{Group 1's result among achievement tests for mathematics}
    \begin{tabular}{ccccccc}
      \hline

                                 &
      \begin{tabular}{c}[Org. A]\\4 M.\end{tabular} &
      \begin{tabular}{c}[Org. A]\\5 M.\end{tabular} &
      \begin{tabular}{c}[Org. B]\\6 M. A\end{tabular} &
      \begin{tabular}{c}[Org. B]\\6 M. B\end{tabular} &
      \begin{tabular}{c}[Org. A]\\7 M.\end{tabular} &
      \begin{tabular}{c}[Org. A]\\8 M.\end{tabular}   \\\hline

      \begin{tabular}{c}[Org. A]\\4 M.\end{tabular} &
                                 &
      $0.82^{\ast\ast}$          &
      $0.67^{\ast\ast}$          &
      $0.58^{\ast\ast}$          &
      $0.77^{\ast\ast}$          &
      $0.67^{\ast\ast}$            \\\hline

      \begin{tabular}{c}[Org. A]\\5 M.\end{tabular} &
                                 &
                                 &
      $0.69^{\ast\ast}$          &
      $0.61^{\ast\ast}$          &
      $0.85^{\ast\ast}$          &
      $0.68^{\ast\ast}$            \\\hline

      \begin{tabular}{c}[Org. B]\\6 M. A\end{tabular} &
                                 &
                                 &
                                 &
      $0.68^{\ast\ast}$          &
      $0.69^{\ast\ast}$          &
      $0.54^{\ast\ast}$            \\\hline

      \begin{tabular}{c}[Org. B]\\6 M. B\end{tabular} &
                                 &
                                 &
                                 &
                                 &
      $0.59^{\ast\ast}$          &
      $0.54^{\ast\ast}$            \\\hline

      \begin{tabular}{c}[Org. A]\\7 M.\end{tabular} &
                                 &
                                 &
                                 &
                                 &
                                 &
      $0.72^{\ast\ast}$            \\\hline

      \begin{tabular}{c}[Org. A]\\8 M.\end{tabular} &
                                 &
                                 &
                                 &
                                 &
                                 &
      \\\hline
    \end{tabular}
    \\Note. $ ^{\ast\ast} p<.01$
    \label{table:table9}
  \end{center}
\end{table}

\begin{table}[htb]
  \begin{center}
    \caption{Group 2's result among achievement tests for national language}
    \begin{tabular}{cccccccc}
      \hline

                                 &
      \begin{tabular}{c}[Org. A]\\5 NL.\end{tabular} &
      \begin{tabular}{c}[Org. B]\\6 NL. A\end{tabular} &
      \begin{tabular}{c}[Org. B]\\6 NL. B\end{tabular} &
      \begin{tabular}{c}[Org. A]\\7 NL.\end{tabular} &
      \begin{tabular}{c}[Org. A]\\8 NL.\end{tabular} &
      \begin{tabular}{c}[Org. B]\\9 NL. A\end{tabular} &
      \begin{tabular}{c}[Org. B]\\9 NL. B\end{tabular}
      \\\hline

      \begin{tabular}{c}[Org. A]\\5 NL.\end{tabular} &
                                 &
      $0.50^{\ast\ast}$          &
      $0.46^{\ast\ast}$          &
      $0.70^{\ast\ast}$          &
      $0.75^{\ast\ast}$          &
      $0.69^{\ast\ast}$          &
      $0.57^{\ast\ast}$            \\\hline

      \begin{tabular}{c}[Org. B]\\6 NL. A\end{tabular} &
                                 &
                                 &
      $0.68^{\ast\ast}$          &
      $0.40^{\ast\ast}$          &
      $0.42^{\ast\ast}$          &
      $0.37^{\ast\ast}$          &
      $0.38^{\ast\ast}$            \\\hline

      \begin{tabular}{c}[Org. B]\\6 NL. B\end{tabular} &
                                 &
                                 &
                                 &
      $0.37^{\ast\ast}$          &
      $0.38^{\ast\ast}$          &
      $0.38^{\ast\ast}$          &
      $0.36^{\ast\ast}$            \\\hline

      \begin{tabular}{c}[Org. A]\\7 NL.\end{tabular} &
                                 &
                                 &
                                 &
                                 &
      $0.72^{\ast\ast}$          &
      $0.68^{\ast\ast}$          &
      $0.53^{\ast\ast}$            \\\hline

      \begin{tabular}{c}[Org. A]\\8 NL.\end{tabular} &
                                 &
                                 &
                                 &
                                 &
                                 &
      $0.70^{\ast\ast}$          &
      $0.62^{\ast\ast}$            \\\hline

      \begin{tabular}{c}[Org. B]\\9 NL. A\end{tabular} &
                                 &
                                 &
                                 &
                                 &
                                 &
                                 &
      $0.60^{\ast\ast}$            \\\hline
      \begin{tabular}{c}[Org. B]\\9 NL. B\end{tabular} &
                                 &
                                 &
                                 &
                                 &
                                 &
                                 &
      \\\hline
    \end{tabular}
    \\Note. $ ^{\ast\ast} p<.01$
    \label{table:table10}
  \end{center}
\end{table}

\begin{table}[htb]
  \begin{center}
    \caption{Group 2's result among achievement tests for mathematics}
    \begin{tabular}{cccccccc}
      \hline

                                 &
      \begin{tabular}{c}[Org. A]\\5 M.\end{tabular} &
      \begin{tabular}{c}[Org. B]\\6 M. A\end{tabular} &
      \begin{tabular}{c}[Org. B]\\6 M. B\end{tabular} &
      \begin{tabular}{c}[Org. A]\\7 M.\end{tabular} &
      \begin{tabular}{c}[Org. A]\\8 M.\end{tabular} &
      \begin{tabular}{c}[Org. B]\\9 M. A\end{tabular} &
      \begin{tabular}{c}[Org. B]\\9 M. B\end{tabular}
      \\\hline

      \begin{tabular}{c}[Org. A]\\5 M.\end{tabular} &
                                 &
      $0.54^{\ast\ast}$          &
      $0.51^{\ast\ast}$          &
      $0.79^{\ast\ast}$          &
      $0.72^{\ast\ast}$          &
      $0.71^{\ast\ast}$          &
      $0.38^{\ast\ast}$            \\\hline

      \begin{tabular}{c}[Org. B]\\6 M. A\end{tabular} &
                                 &
                                 &
      $0.69^{\ast\ast}$          &
      $0.50^{\ast\ast}$          &
      $0.44^{\ast\ast}$          &
      $0.46^{\ast\ast}$          &
      $0.40^{\ast\ast}$            \\\hline

      \begin{tabular}{c}[Org. B]\\6 M. B\end{tabular} &
                                 &
                                 &
                                 &
      $0.43^{\ast\ast}$          &
      $0.42^{\ast\ast}$          &
      $0.37^{\ast\ast}$          &
      $0.41^{\ast\ast}$            \\\hline

      \begin{tabular}{c}[Org. A]\\7 M.\end{tabular} &
                                 &
                                 &
                                 &
                                 &
      $0.78^{\ast\ast}$          &
      $0.80^{\ast\ast}$          &
      $0.74^{\ast\ast}$            \\\hline

      \begin{tabular}{c}[Org. A]\\8 M.\end{tabular} &
                                 &
                                 &
                                 &
                                 &
                                 &
      $0.85^{\ast\ast}$          &
      $0.77^{\ast\ast}$            \\\hline

      \begin{tabular}{c}[Org. B]\\9 M. A\end{tabular} &
                                 &
                                 &
                                 &
                                 &
                                 &
                                 &
      $0.82^{\ast\ast}$            \\\hline
      \begin{tabular}{c}[Org. B]\\9 M. B\end{tabular} &
                                 &
                                 &
                                 &
                                 &
                                 &
                                 &
      \\\hline
    \end{tabular}
    \\Note. $ ^{\ast\ast} p<.01$
    \label{table:table11}
  \end{center}
\end{table}

\clearpage
\subsection{Time-series clustering step}
In this step, we targeted the mathematics achievement test trends. The Data Screening step excluded some achievement tests. Thus, we used [Org. A] 5 M., [Org. A] 7 M., and [Org. A] 8 M., as both group 1 and group 2 included these tests.

Our approach had to consider both the timeline shape and value. A combination of DTW and k-means clustering is one of the methods appropriate for such a purpose. However, the input data set included data from only three achievement tests; thus, the input data set was unsuitable for DTW. Therefore, we translated the input data set to a vector represented by formula (2) and clustered the vectors with k-means clustering.

\begin{equation}
  \left(\begin{array}{c}x_{1,i}\\x_{2,i}\\x_{3,i}\\x_{4,i}\\x_{5,i}\\x_{6,i}\end{array}\right)=\left(\begin{array}{c}T.\,{\rm of\,[Org.A]\,5\,M.}_i\\T.\,{\rm of\,[Org.A]\,7\,M.}_i\\T.\,{\rm of\,[Org.A]\,8\,M.}_i\\T.\,{\rm of\,[Org.A]\,8\,M.}_i-T.\,{\rm of\,[Org.A]\,5\,M.}_i\\T.\,{\rm of\,[Org.A]\,8\,M.}_i-T.\,{\rm of\,[Org.A]\,7\,M.}_i\\T.\,{\rm of\,[Org.A]\,7\,M.}_i-T.\,{\rm of\,[Org.A]\,5\,M.}_i\end{array}\right).
\end{equation}

We used $x_{1,i}$, $x_{2,i}$, and $x_{3,i}$ to consider the value of the student achievement data, and $x_{4,i}$, $x_{5,i}$, and $x_{6,i}$ to consider the shape of the student achievement data. Then, we clustered the vectors with Euclidean distance and k-means, as we set the number of clusters as four. Additionally, when we use another number of clusters, we may extract another variation factor.

\begin{figure}[htbp]
  \centering
  \includegraphics[width=10cm]{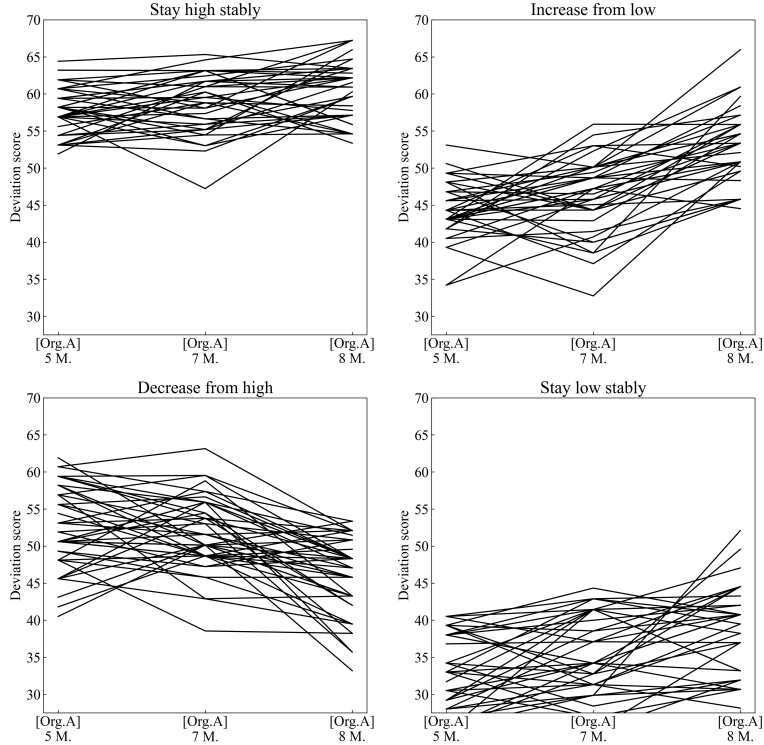}
  \caption{Results of group 1 clustering}
  \label{fig:fig3}
\end{figure}

\begin{figure}[htbp]
  \centering
  \includegraphics[width=10cm]{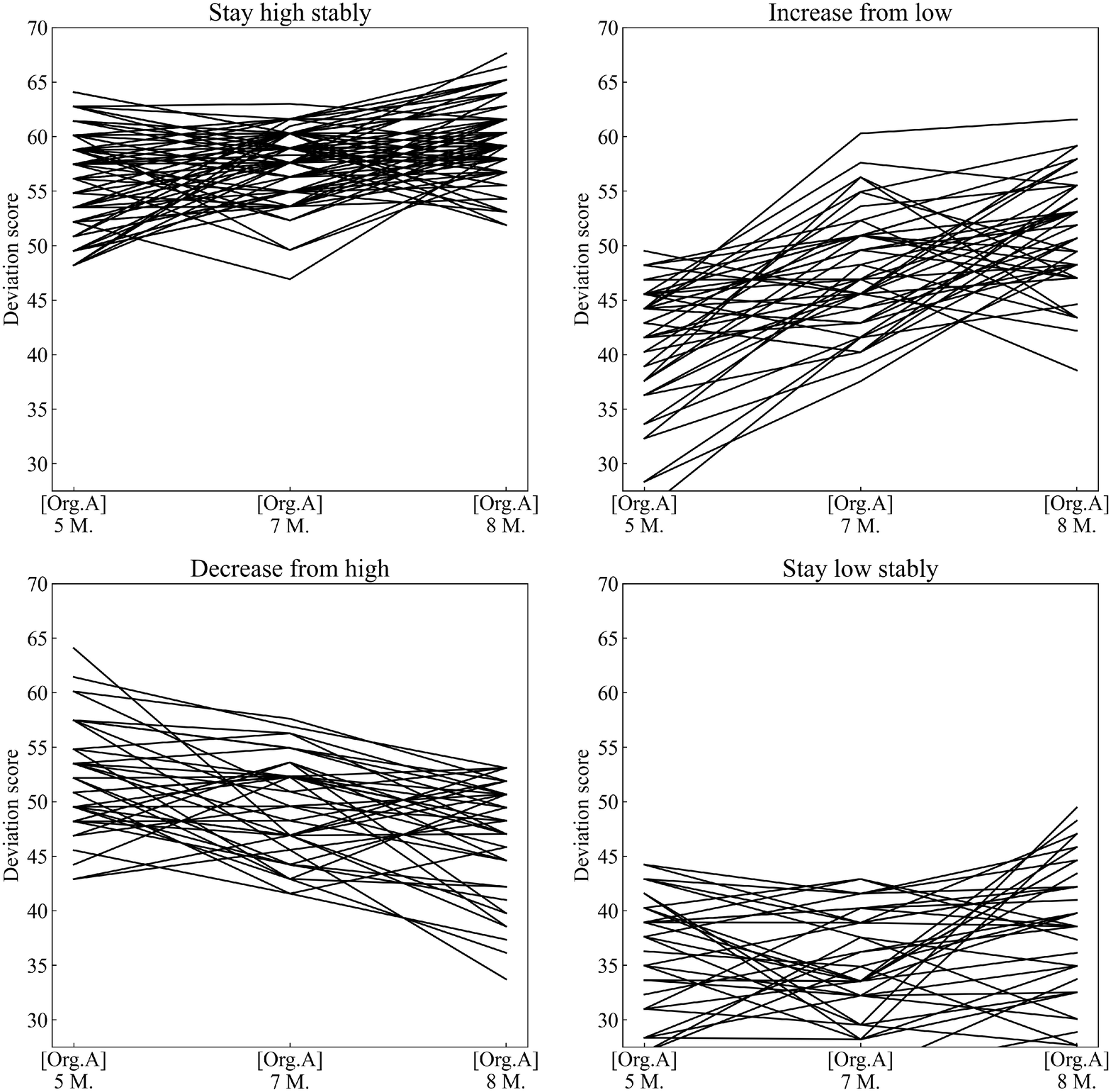}
  \caption{Results of group 2 clustering}
  \label{fig:fig4}
\end{figure}

Figure \ref{fig:fig3}, Figure \ref{fig:fig4}, Table \ref{table:table12}, and Table \ref{table:table13} illustrate the results of group 1 clustering and group 2 clustering. We named the groups of results after their shapes and values: ``stay high stably,'' ``stay low stably,'' ``increase from low,'' and ``decrease from high.'' As a result, we clustered the achievement tests of both groups 1 and 2 into the same four clusters.

\begin{table}[htbp]
  \centering
  \caption{Fundamental statistics of group 1 clustering}
  \begin{tabular}{cccccccc} \hline
                       &                    & \multicolumn{2}{c}{$T$. of [Org. A] 5 M.} & \multicolumn{2}{c}{$T$. of [Org. A] 7 M.} & \multicolumn{2}{c}{$T$. of [Org. A] 8 M.}                       \\ \hline
    Cluster            & Number of Students & Avg.                                        & S.D.                                        & Avg.                                        & S.D. & Avg.  & S.D. \\ \hline
    Stay high stably   & 39                 & 57.93                                       & 3.26                                        & 58.86                                       & 3.99 & 60.33 & 3.91 \\ \hline
    Increase from low  & 40                 & 44.41                                       & 3.97                                        & 46.34                                       & 4.99 & 52.70 & 4.74 \\ \hline
    Decrease from high & 45                 & 52.34                                       & 5.38                                        & 51.66                                       & 4.86 & 45.97 & 5.13 \\ \hline
    Stay low stably    & 37                 & 32.80                                       & 5.27                                        & 35.15                                       & 5.55 & 38.39 & 6.08 \\ \hline
  \end{tabular}
  \label{table:table12}
\end{table}

\begin{table}[htbp]
  \centering
  \caption{Fundamental statistics of group 2 clustering}
  \begin{tabular}{cccccccc} \hline
                       &                    & \multicolumn{2}{c}{$T$. of [Org. A] 5 M.} & \multicolumn{2}{c}{$T$. of [Org. A] 7 M.} & \multicolumn{2}{c}{$T$. of [Org. A] 8 M.}                       \\ \hline
    Cluster            & Number of Students & Avg.                                        & S.D.                                        & Avg.                                        & S.D. & Avg.  & S.D. \\ \hline
    Stay high stably   & 80                 & 55.98                                       & 4.42                                        & 57.59                                       & 3.14 & 59.72 & 3.51 \\ \hline
    Increase from low  & 48                 & 41.21                                       & 5.60                                        & 47.58                                       & 5.25 & 51.27 & 4.97 \\ \hline
    Decrease from high & 40                 & 52.03                                       & 5.02                                        & 49.57                                       & 4.59 & 46.70 & 5.27 \\ \hline
    Stay low stably    & 36                 & 35.69                                       & 5.68                                        & 34.92                                       & 5.03 & 38.82 & 6.12 \\ \hline
  \end{tabular}
  \label{table:table13}
\end{table}

\clearpage
\subsection{Causal inference step}
Causal inference is the identification of the cause of a phenomenon. In this paper, we used multivariate logistic regression because of its popularity.

We compared ``stay high stably'' and ``decrease from high,'' and ``stay low stably'' and ``increase from low.'' The target variable of multivariate logistic regression was the cluster type ``stay high stably,'' which was 1; ``decrease from high,'' which was 0; ``increase from low,'' which was 1; and ``decrease from high,'' which was 0. The explanatory variables were the score of [Org. A] 5 NL to extract the national language's factors affecting the long-term trends of mathematics. The score represented correct and incorrect as 1 and 0, respectively, in each item. We used [Org. A] 5 NL. as the evidence variable because it was the starting point of the long-term data set. Tables \ref{table:table14} and \ref{table:table15} illustrate the items of [Org. A] 5 NL. in 2014 and 2015 and their topics. We selected the explanatory variables by the variable reduction method and performed a stepwise removal to eliminate the items with variance inflation factors higher than 10.

Table \ref{table:table16} illustrates the multivariate logistic regression results of groups 1 and 2. The R-squared of group 1 was .15, and the $p$-value was .18. The item with a $p$-value of less than .10 was Item ID 2014-24 ``interpret the information of the text and make a supplementary statement,'' and the coefficient was 1.9788$^\ast$. The R-squared of group 2 was .29, and the $p$-value was .02. The items with $p$-values of less than .10 were as follows: Item ID 2015-3 ``collaborate with others considering the others' ideas,'' which had a coefficient of 2.32$^\ast$; Item ID 2015-12 ``interpret Japanese grammar,'' which had a coefficient of 1.69$^\ast$; Item ID 2015-13 ``interpret Japanese grammar,'' which had a coefficient of 2.47$^{\ast\ast}$; Item ID 2015-22 ``read the text considering the connection between paragraphs,'' which had a coefficient of 1.14$^\dagger$; Item ID 2015-24 ``interpret the information of the text and make a supplementary statement,'' which had a coefficient of 2.85$^{\ast\ast}$; Item ID 2015-26 ``summarize the content of the interview and the impressions of the interviewer considering the purpose,'' which had a coefficient of -1.51$^\dagger$; and  $^\dagger p<.10$; $^\ast p<.05$; $^{\ast\ast}p<.01$.

Table \ref{table:table17} illustrates the multivariate logistic regression results of groups 1 and 2. The R-squared of group 1 was .28, and the $p$-value was .16. The items with $p$-values of less than .10 were as follows: Item ID 2014-17 ``read a character's feelings,'' which had a coefficient of -1.99$^\ast$; Item ID 2014-19 ``read a character's feelings depending on the purpose,'' which had a coefficient of 1.24$^\dagger$; Item ID 2014-24 ``interpret the information of the text and make a supplementary statement,'' which had a coefficient of 3.01$^\dagger$; and Item ID 2014-25 ``write a sentence within a word limit,'' which had a coefficient of 2.53$^\ast$.

The R-squared of group 2 was .52, and the $p$-value was .0002. The items with $p$-values of less than .10 were as follows: Item ID 2015-7 ``read a kanji character,'' which had a coefficient of -3.82$^\ast$; Item ID 2015-9 ``write a kanji character,'' which had a coefficient of 2.83$^\dagger$; Item ID 2015-13 ``interpret Japanese grammar,'' which had a coefficient of -2.77$^\dagger$; Item ID 2015-18 ``read the situation of the text,'' which had a coefficient of 2.26$^\dagger$; Item ID 2015-19 ``read the text depending on the purpose,'' which had a coefficient of 2.46$^\dagger$; Item ID 2015-21 ``read the text precisely,'' which had a coefficient of -3.03$^\ast$; Item ID 2015-25 ``write a sentence within a word limit,'' which had a coefficient of 2.72$^\dagger$; Item ID 2015-27 ``summarize the content of the interview considering the purpose,'' which had a coefficient of 2.62$^\dagger$; and $^\dagger p<.10$; $^\ast p<.05$; $^{\ast\ast}p<.01$.

\begin{table}[htbp]
  \centering
  \caption{Item of [Org. A] 5 NL. in 2014}
  \begin{tabular}{cc} \hline
    Item ID & Topic                                                                                                 \\ \hline
    2014-1  & listen to the conversation considering the central theme                                              \\ \hline
    2014-2  & listen to the conversation considering the central theme                                              \\ \hline
    2014-3  & collaborate with others considering the others' ideas                                                 \\ \hline
    2014-4  & read a kanji character                                                                                \\ \hline
    2014-5  & read a kanji character                                                                                \\ \hline
    2014-6  & read a kanji character                                                                                \\ \hline
    2014-7  & read a kanji character                                                                                \\ \hline
    2014-8  & write a kanji character                                                                               \\ \hline
    2014-9  & write a kanji character                                                                               \\ \hline
    2014-10 & write a kanji character                                                                               \\ \hline
    2014-11 & write a kanji character                                                                               \\ \hline
    2014-12 & interpret Japanese grammar                                                                            \\ \hline
    2014-13 & interpret Japanese grammar                                                                            \\ \hline
    2014-14 & interpret Japanese grammar                                                                            \\ \hline
    2014-15 & use a dictionary                                                                                      \\ \hline
    2014-16 & read a character's feelings                                                                           \\ \hline
    2014-17 & read a character's feelings                                                                           \\ \hline
    2014-18 & read the situation of the text                                                                        \\ \hline
    2014-19 & read a character's feelings depending on the purpose                                                  \\ \hline
    2014-20 & read the text considering the connection between sentences                                            \\ \hline
    2014-21 & read the text depending on the purpose                                                                \\ \hline
    2014-22 & read the text considering the connection between paragraphs                                           \\ \hline
    2014-23 & interpret the information of the text and modify the text                                             \\ \hline
    2014-24 & interpret the information of the text and make a supplementary statement                              \\ \hline
    2014-25 & write a sentence within a word limit                                                                  \\ \hline
    2014-26 & use a paragraph structure                                                                             \\ \hline
    2014-27 & summarize the content of the interview considering the purpose                                        \\ \hline
    2014-28 & summarize the content of the interview and the impressions of the interviewer considering the purpose \\ \hline
  \end{tabular}
  \label{table:table14}
\end{table}

\begin{table}[htbp]
  \centering
  \caption{Item of [Org. A] 5 NL. in 2015}
  \begin{tabular}{cc} \hline
    Item ID & Topic                                                                                                 \\ \hline
    2015-1  & listen to the conversation considering the main theme                                                 \\ \hline
    2015-2  & listen to the conversation considering the ingenuity of the speaker                                   \\ \hline
    2015-3  & collaborate with others considering the others' ideas                                                 \\ \hline
    2015-4  & read a kanji character                                                                                \\ \hline
    2015-5  & read a kanji character                                                                                \\ \hline
    2015-6  & read a kanji character                                                                                \\ \hline
    2015-7  & read a kanji character                                                                                \\ \hline
    2015-8  & write a kanji character                                                                               \\ \hline
    2015-9  & write a kanji character                                                                               \\ \hline
    2015-10 & write a kanji character                                                                               \\ \hline
    2015-11 & write a kanji character                                                                               \\ \hline
    2015-12 & interpret Japanese grammar                                                                            \\ \hline
    2015-13 & interpret Japanese grammar                                                                            \\ \hline
    2015-14 & interpret Japanese grammar                                                                            \\ \hline
    2015-15 & use a dictionary                                                                                      \\ \hline
    2015-16 & read a character's feelings                                                                           \\ \hline
    2015-17 & read a character's feelings                                                                           \\ \hline
    2015-18 & read the situation of the text                                                                        \\ \hline
    2015-19 & read a character's feelings depending on the purpose                                                  \\ \hline
    2015-20 & read the text precisely                                                                               \\ \hline
    2015-21 & read the text precisely                                                                               \\ \hline
    2015-22 & read the text considering the connection between paragraphs                                           \\ \hline
    2015-23 & interpret the information of the text and select the picture corresponding to it                      \\ \hline
    2015-24 & interpret the information of the text and make a supplementary statement                              \\ \hline
    2015-25 & write a sentence within a word limit                                                                  \\ \hline
    2015-26 & use a paragraph structure                                                                             \\ \hline
    2015-27 & summarize the content of the interview considering the purpose                                        \\ \hline
    2015-28 & summarize the content of the interview and the impressions of the interviewer considering the purpose \\ \hline
  \end{tabular}
  \label{table:table15}
\end{table}

\begin{table}[htbp]
  \centering
  \caption{Result between “stay high stably” and “decrease from high”}
  \begin{tabular}{cccccc} \hline
    \multicolumn{3}{c}{Group 1} & \multicolumn{3}{c}{Group 2}                                              \\ \hline
    Item ID                       & coef.                         & $p$   & Item ID    & coef.             & $p$   \\ \hline
    [2014G5]1                     & 0.12                          & .85 & [2015G5]3  & 2.32$^\ast$       & .03 \\ \hline
    [2014G5]3                     & 0.40                          & .47 & [2015G5]8  & -0.40             & .54 \\ \hline
    [2014G5]8                     & -0.89                         & .17 & [2015G5]9  & -0.63             & .42 \\ \hline
    [2014G5]9                     & 0.78                          & .11 & [2015G5]10 & -1.02             & .23 \\ \hline
    [2014G5]10                    & -0.30                         & .58 & [2015G5]11 & -1.04             & .24 \\ \hline
    [2014G5]11                    & -0.03                         & .96 & [2015G5]12 & 1.69$^\ast$       & .04 \\ \hline
    [2014G5]13                    & 0.76                          & .12 & [2015G5]13 & 2.47$^{\ast\ast}$ & .00 \\ \hline
    [2014G5]15                    & -0.69                         & .14 & [2015G5]15 & 0.56              & .41 \\ \hline
    [2014G5]16                    & 0.05                          & .93 & [2015G5]17 & -1.30             & .18 \\ \hline
    [2014G5]17                    & 0.49                          & .45 & [2015G5]18 & 0.46              & .46 \\ \hline
    [2014G5]18                    & -0.17                         & .74 & [2015G5]19 & -0.91             & .31 \\ \hline
    [2014G5]20                    & 0.55                          & .25 & [2015G5]20 & -0.85             & .32 \\ \hline
    [2014G5]21                    & -0.08                         & .88 & [2015G5]21 & -0.42             & .54 \\ \hline
    [2014G5]22                    & -0.04                         & .94 & [2015G5]22 & 1.14$^\dagger$    & .08 \\ \hline
    [2014G5]24                    & 1.98$^\ast$                   & .02 & [2015G5]23 & -0.39             & .58 \\ \hline
    [2014G5]25                    & 0.33                          & .60 & [2015G5]24 & 2.85$^{\ast\ast}$ & .00 \\ \hline
    [2014G5]26                    & 0.13                          & .80 & [2015G5]25 & 0.58              & .57 \\ \hline
    [2014G5]27                    & 0.18                          & .76 & [2015G5]26 & -1.51$^\dagger$   & .09 \\ \hline
    [2014G5]28                    & 0.14                          & .80 & [2015G5]27 & 0.77              & .33 \\ \hline
                                  &                               &     & [2015G5]28 & -0.61             & .41 \\ \hline
  \end{tabular}
  \\Note. $^\dagger p<.10$; $^\ast p<.05$; $^{\ast\ast}p<.01$
  \label{table:table16}
\end{table}

\begin{table}[htbp]
  \centering
  \caption{Result between “stay low stably” and “increase from low”}
  \begin{tabular}{cccccc} \hline
    \multicolumn{3}{c}{Group 1} & \multicolumn{3}{c}{Group 2}                                            \\ \hline
    Item ID                       & coef.                         & $p$   & Item ID    & coef.             & $p$   \\ \hline
    [2014G5]1                     & 0.82                          & .29 & [2015G5]2  & -0.55           & .61 \\ \hline
    [2014G5]2                     & -1.15                         & .13 & [2015G5]3  & -1.95           & .26 \\ \hline
    [2014G5]3                     & 0.32                          & .83 & [2015G5]7  & -3.82$^\ast$    & .03 \\ \hline
    [2014G5]6                     & -0.24                         & .77 & [2015G5]8  & 0.81            & .50 \\ \hline
    [2014G5]7                     & 0.16                          & .86 & [2015G5]9  & 2.83$^\dagger$  & .05 \\ \hline
    [2014G5]8                     & 1.06                          & .21 & [2015G5]10 & 1.24            & .36 \\ \hline
    [2014G5]9                     & 0.23                          & .77 & [2015G5]11 & 0.89            & .53 \\ \hline
    [2014G5]10                    & -0.21                         & .77 & [2015G5]12 & -0.57           & .50 \\ \hline
    [2014G5]11                    & -0.71                         & .35 & [2015G5]13 & -2.77$^\dagger$ & .06 \\ \hline
    [2014G5]12                    & 0.12                          & .86 & [2015G5]14 & -1.33           & .31 \\ \hline
    [2014G5]13                    & -0.14                         & .89 & [2015G5]15 & -2.28           & .14 \\ \hline
    [2014G5]14                    & 0.004                         & .99 & [2015G5]16 & 0.57            & .67 \\ \hline
    [2014G5]15                    & 0.97                          & .13 & [2015G5]17 & -0.81           & .45 \\ \hline
    [2014G5]16                    & -0.17                         & .83 & [2015G5]18 & 2.26$^\dagger$  & .09 \\ \hline
    [2014G5]17                    & -2.00$^\ast$                  & .02 & [2015G5]19 & 2.46$^\dagger$  & .08 \\ \hline
    [2014G5]18                    & -0.09                         & .91 & [2015G5]20 & 2.69            & .06 \\ \hline
    [2014G5]19                    & 1.24$^\dagger$                & .09 & [2015G5]21 & -3.03$^\ast$    & .07 \\ \hline
    [2014G5]20                    & 0.21                          & .78 & [2015G5]22 & -1.53           & .18 \\ \hline
    [2014G5]21                    & -0.04                         & .95 & [2015G5]23 & 0.90            & .43 \\ \hline
    [2014G5]22                    & 0.04                          & .96 & [2015G5]24 & 4.92            & .12 \\ \hline
    [2014G5]23                    & -0.09                         & .90 & [2015G5]25 & 2.72$^\dagger$  & .08 \\ \hline
    [2014G5]24                    & 3.01$^\dagger$                & .07 & [2015G5]26 & 0.97            & .50 \\ \hline
    [2014G5]25                    & 2.53$^\ast$                   & .01 & [2015G5]27 & 2.62$^\dagger$  & .07 \\ \hline
    [2014G5]26                    & 1.30                          & .43 & [2015G5]28 & -1.96           & .18 \\ \hline
    [2014G5]27                    & -0.84                         & .44 &            &                 &     \\ \hline
    [2014G5]28                    & -1.26                         & .44 &            &                 &     \\ \hline
  \end{tabular}
  \\Note. $^\dagger p<.10$; $^\ast p<.05$; $^{\ast\ast}p<.01$
  \label{table:table17}
\end{table}

\clearpage
\subsection{Summary}
We applied a multivariate logistic regression to the achievement test data to extract the national language factors affecting the long-term mathematics trends.

First, we analyzed ``stay high stably'' and ``decrease from high'' to extract the variation factors for why the score decreased from high or stayed high stably. The results demonstrate that the common variation factor between groups 1 and 2 was ``interpret the information of the text and make a supplementary statement.'' For this item, students needed to interpret problematic texts, represent their ideas as text, pay attention to the context, and choose appropriate vocabulary. These results imply that this ability is important in keeping mathematics scores high. This ability seems to be related to the number of words and vocabulary students know \cite{Mullis2011}.

Next, we analyzed ``stay low stably'' and ``increase from low'' to extract the variation factors for why the score increased from low or why it stayed low stably. The results demonstrate that the common variation factors between groups 1 and 2 were ``write a sentence within a word limit'' and ``read a character's feelings depending on the purpose.'' The first item tests the ability to write a sentence in the specified number of words, questioning the number of words and vocabulary the students possess \cite{Mullis2011}. The results indicate the importance of this ability in increasing a low mathematics score. The second item tests the ability to guess the character's state of mind according to a purpose. To the best of our knowledge, no study has examined this factor. This result implies a unique relationship between mathematics skills and the ability to find information and interpret it, considering purpose and circumstances.

Some R-squared were small (e.g., .15, .16), and some $p$-values were over .10. However, this model estimates four years' worth of trends of mathematics scores based on the national language scores of the fifth grade. While this task is very challenging, we believe its results can help formulate the hypothesis that certain skills could impact future skills affecting other subjects.

\clearpage
\section{Evaluation Experiment 2}
In this experiment, we skipped the Data screening step to validate its effectiveness. We used [Org. A] 5 M., [Org. B] 6 M. A, [Org. B] 6 M. B, [Org. A] 7 M., and [Org. A] 8 M. because both groups 1 and 2 included these tests. Then, we translated the long-term data sets into a vector, which was represented by formula (3); C.: correct answer ratio.

\begin{equation}
  \left(\begin{array}{c}x_{1,i}\\x_{2,i}\\x_{3,i}\\x_{4,i}\\x_{5,i}\\x_{6,i}\\x_{7,i}\\x_{8,i}\\x_{9,i}\\x_{10,i}\\x_{11,i}\\x_{12,i}\\x_{13,i}\\x_{14,i}\\x_{15,i}\end{array}\right)=\left(\begin{array}{c}
    C.\,{\rm of\,[Org.A]\,5\,M.}_i                                      \\
    C.\,{\rm of\,[Org.A]\,6\,M.A.}_i                                    \\
    C.\,{\rm of\,[Org.A]\,6\,M.B.}_i                                    \\
    C.\,{\rm of\,[Org.A]\,7\,M.}_i                                      \\
    C.\,{\rm of\,[Org.A]\,8\,M.}_i                                      \\
    C.\,{\rm of\,[Org.A]\,8\,M.}_i - C.\,{\rm of\,[Org.A]\,5\,M.}_i     \\
    C.\,{\rm of\,[Org.A]\,8\,M.}_i - C.\,{\rm of\,[Org.A]\,6\,M.A.}_i   \\
    C.\,{\rm of\,[Org.A]\,8\,M.}_i - C.\,{\rm of\,[Org.A]\,6\,M.B.}_i   \\
    C.\,{\rm of\,[Org.A]\,8\,M.}_i - C.\,{\rm of\,[Org.A]\,7\,M.}_i     \\
    C.\,{\rm of\,[Org.A]\,7\,M.}_i - C.\,{\rm of\,[Org.A]\,5\,M.}_i     \\
    C.\,{\rm of\,[Org.A]\,7\,M.}_i - C.\,{\rm of\,[Org.A]\,6\,M.A.}_i   \\
    C.\,{\rm of\,[Org.A]\,7\,M.}_i - C.\,{\rm of\,[Org.A]\,6\,M.B.}_i   \\
    C.\,{\rm of\,[Org.A]\,6\,M.B.}_i - C.\,{\rm of\,[Org.A]\,5\,M.}_i   \\
    C.\,{\rm of\,[Org.A]\,6\,M.B.}_i - C.\,{\rm of\,[Org.A]\,6\,M.A.}_i \\
    C.\,{\rm of\,[Org.A]\,6\,M.A.}_i - C.\,{\rm of\,[Org.A]\,5\,M.}_i   \\
  \end{array}\right).
\end{equation}

\begin{figure}[htbp]
  \centering
  \includegraphics[width=10cm]{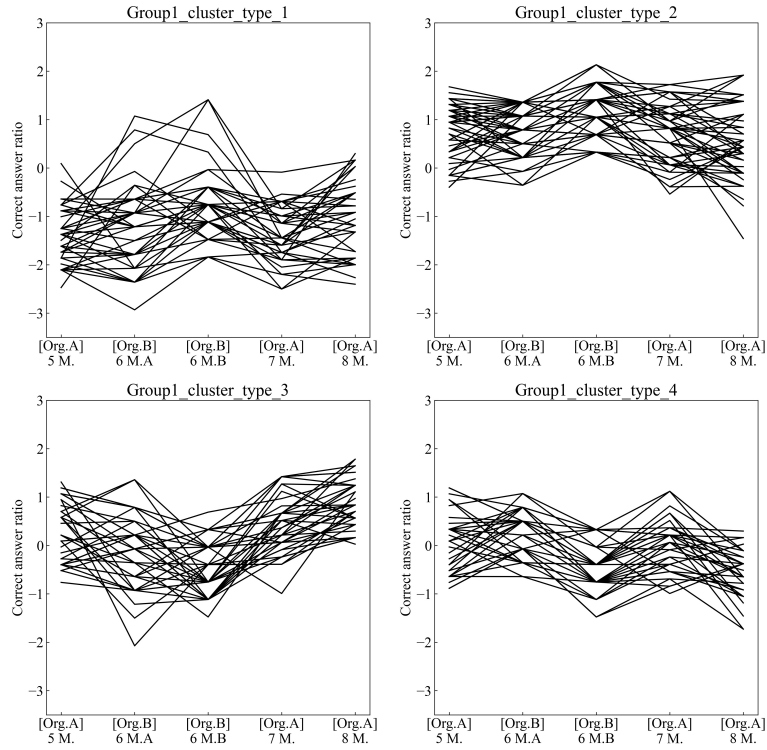}
  \caption{Group 1's results in experiment 2; cluster number as 4}
  \label{fig:fig5}
\end{figure}

We changed the number of clusters from two to six. Figure \ref{fig:fig5} illustrates group 1's results as we set the cluster number as 4, and Table \ref{table:table18} illustrates the fundamental statistics of each test. The other results are in the Appendices (Figures \ref{fig:fig6}, \ref{fig:fig7}, \ref{fig:fig8}, \ref{fig:fig9}, \ref{fig:fig10}, \ref{fig:fig11}, \ref{fig:fig12}, \ref{fig:fig13}, \ref{fig:fig14}, \ref{fig:fig15}).

In Figure \ref{fig:fig5}, all clusters decreased at [Org. B] 6 M. B., which had a low correlation coefficient, and the shapes of the clusters were not easily interpretable. Moreover, in the Appendices, the other clustering results were scattered, or the shapes of the results changed at the tests, which had low correlation coefficients. According to these results, when the long-term data include different evaluation criteria, we cannot cluster them into interpretable groups.

\clearpage

\begin{table}[htbp]
  \centering
  \caption{Fundamental statistics of answer ratio in experiment 2}
  \begin{tabular}{cccccc} \hline
                    &
    [Org. A] 5 M.   &
    [Org. B] 6 M. A &
    [Org. B] 6 M. B &
    [Org. A] 7 M.   &
    [Org. A] 8 M.
    \\\hline
    Avg.            & 0.64 & 0.70 & 0.39 & 0.65 & 0.57 \\ \hline
    S.D.            & 0.22 & 0.22 & 0.21 & 0.57 & 0.21 \\ \hline
  \end{tabular}
  \label{table:table18}
\end{table}

\clearpage
\subsection{Discussion}
Evaluation experiment 1 demonstrates that our approach extracted coherence data, clustered the long-term data into interpretable groups, and extracted candidate factors affecting academic ability across subjects.

The most crucial step was the data screening step. In this step, our approach focused on the correlation coefficients between tests of two consecutive times; then, our approach excluded the test data that had low correlation coefficients with other tests. It is difficult to believe that many students' scores suddenly change at the same time, and it is easier to believe that the evaluation criteria of the tests changed. Surprisingly, experiment 1 revealed that even a test designed by the same organization, organization B, lacked coherence. We evaluated experiment 2 without the data screening step. The clustering results were scattered, or the shapes of the results changed at the tests, which showed low correlation coefficients.

We now consider the time series clustering step. Although our study adopted only one of the time series clustering methods, our approach can adopt other methods. First, our approach can adopt Aghabozorgi, Shirkhorshidi, and Wah's \cite{Aghabozorgi2015} two types of utilization pattern discovery of time series data. One is to discover patterns that frequently appear in the time series data \cite{Chiu2003}. The evaluation experiment is categorized as this type as our approach clustered the achievement tests of both groups 1 and 2 into the same four clusters. The other is to discover patterns that occur outstandingly in time series data \cite{Chan2005,Wei2005}. When we used appropriate achievement test data, our approach could elucidate outstanding data, such as data indicating a student who suddenly increased their score.

Further, our approach can adopt the three types of clustering analyses for time series data \cite{Aghabozorgi2015,Keogh2003}: Whole-time series clustering for clustering individual time series based on their similarity; subsequence clustering for clustering subsets of each time series, the subsets are extracted via a sliding window; and Timepoint clustering for clustering time points based on a combination of their temporal proximity of time points and the similarity of the corresponding values.

The evaluation experiment utilized the whole time series clustering pattern: the student achievement data were translated into a vector and adopted k-means clustering because they were not rich. When we used rich achievement test data, it was possible to adopt subsequent and time point clustering.

Further, our approach adopted the three methods from Aghabozorgi, Shirkhorshidi, and Wah \cite{Aghabozorgi2015}: the shape-based method (raw-data-based method), feature-based method, and model-based method. The shape-based method matches the shapes of the two-time series by a non-linear stretching and contraction of the time axes. Then, conventional clustering methods are applied by modifying distance/similarity measures for time series data. The shape-based method converts the raw time series into a feature vector for a lower dimension. After that, conventional clustering methods are applied to the extracted feature vectors. The model-based method transforms the raw time series into model parameters. Finally, a suitable model distance and a clustering algorithm are applied to the extracted model parameters.

The evaluation experiment utilized the shape-based method (raw-data-based method). When we use rich achievement test data, our approach can adopt the feature-based method. Further, when we can assume the student learning model and use rich achievement test data, our approach can adopt a model-based method.

\clearpage
\section{Conclusion}
We proposed a novel approach to extract candidate factors affecting the long-term trends of students' abilities across subjects. Then, we conducted evaluation experiments with student achievement data from five public elementary schools and four public junior high schools. The results demonstrate that our approach extracted coherence data series of student achievement tests, clustered the long-term data into interpretable groups, and extracted the long-term mutual interaction across subjects. Surprisingly, the experiment revealed that even tests designed by the same organization lack coherence. In conclusion, with our approach, we were able to formulate the hypotheses illustrated below regarding what affected academic achievement across subjects and grades.

First hypothesis: the ability to write a sentence in the specified number of words, to question the number of words and the degree of vocabulary students possess are important factors in retaining high mathematics scores.

Second hypothesis: the ability to write a sentence in the specified number of words, to question the number of words and the degree of vocabulary students possess are important factors in increasing low mathematics scores.

Third hypothesis: the ability to guess a character's state of mind according to a purpose is an important factor in increasing low mathematics scores.

To the best of our knowledge, no study has examined the third hypothesis. This result implies a unique relationship between mathematics skills and the ability to find information and interpret it, considering purpose and circumstances. We believe that our novel approach can help schoolteachers and educational policymakers extract candidate factors for educational policymaking.

Our work is not without limitations. The causal inference step adopted a multivariate logistic regression, which is a statistical inference \cite{Han2011}. To conclude that these results cause the student achievement data change, we would need to adopt a quantity analysis and an effect size analysis. For instance, we might ask about when schools introduce interventions related to students' ability to interpret a problematic text and represent their ideas as text while paying attention to the context of the text, which are abilities associated with high mathematics scores, how many students can improve their mathematics scores.

Our hypotheses must be confirmed through practice in schools. We are currently working on this, but this will take time. Nevertheless, we believe that our approach can identify students in need of help early on and identify focus topics for teachers.

\clearpage
\bibliographystyle{unsrt}
\bibliography{references}

\begin{thebibliography}{10}

\bibitem{Hwang2018}
Sophia~HJ Hwang and Elise Cappella.
\newblock Rethinking early elementary grade retention: Examining long-term
  academic and psychosocial outcomes.
\newblock {\em Journal of Research on Educational Effectiveness},
  11(4):559--587, 2018.

\bibitem{Vaughn2008}
Sharon Vaughn, Paul~T Cirino, Tammy Tolar, Jack~M Fletcher, Elsa
  Cardenas-Hagan, Coleen~D Carlson, and David~J Francis.
\newblock Long-term follow-up of spanish and english interventions for
  first-grade english language learners at risk for reading problems.
\newblock {\em Journal of Research on Educational Effectiveness},
  1(3):179--214, 2008.

\bibitem{Mullis2011}
Ina~VS Mullis, Michael~O Martin, and Pierre Foy.
\newblock The impact of reading ability on timss mathematics and science
  achievement at the fourth grade: An analysis by item reading demands.
\newblock {\em TIMSS and PIRLS}, pages 67--108, 2011.

\bibitem{Merki2017}
Katharina~Maag Merki and Britta Oerke.
\newblock Long-term effects of the implementation of state-wide exit exams: a
  multilevel regression analysis of mediation effects of teaching practices on
  students’ motivational orientations.
\newblock {\em Educational Assessment, Evaluation and Accountability},
  29(1):23--54, 2017.

\bibitem{Droop2016}
Mienke Droop, Willy van Els{\"a}cker, Marinus~JM Voeten, and Ludo Verhoeven.
\newblock Long-term effects of strategic reading instruction in the
  intermediate elementary grades.
\newblock {\em Journal of Research on Educational Effectiveness}, 9(1):77--102,
  2016.

\bibitem{Watts2017}
Tyler~W Watts, Douglas~H Clements, Julie Sarama, Christopher~B Wolfe,
  Mary~Elaine Spitler, and Drew~H Bailey.
\newblock Does early mathematics intervention change the processes underlying
  children's learning?
\newblock {\em Journal of Research on Educational Effectiveness},
  10(1):96--115, 2017.

\bibitem{Rousseau2006}
Denise~M Rousseau.
\newblock Is there such a thing as “evidence-based management”?
\newblock {\em Academy of management review}, 31(2):256--269, 2006.

\bibitem{Cawley1998}
John Cawley, James~Joseph Heckman, and Edward Vytlacil.
\newblock Cognitive ability and the rising return to education, 1998.

\bibitem{Cunha2009}
Flavio Cunha and James~J Heckman.
\newblock The economics and psychology of inequality and human development.
\newblock {\em Journal of the European Economic Association}, 7(2-3):320--364,
  2009.

\bibitem{Aghabozorgi2015}
Saeed Aghabozorgi, Ali~Seyed Shirkhorshidi, and Teh~Ying Wah.
\newblock Time-series clustering--a decade review.
\newblock {\em Information Systems}, 53:16--38, 2015.

\bibitem{Liao2005}
T~Warren Liao.
\newblock Clustering of time series data―a survey.
\newblock {\em Pattern recognition}, 38(11):1857--1874, 2005.

\bibitem{Kaufman2001}
Robert~R Kaufman and Alex Segura-Ubiergo.
\newblock Globalization, domestic politics, and social spending in latin
  america: a time-series cross-section analysis, 1973--97.
\newblock {\em World politics}, 53(4):553--587, 2001.

\bibitem{Loening2002}
Josef~Ludger Loening.
\newblock The impact of education on economic growth in guatemala: a
  time-series analysis applying an error-correction methodology.
\newblock {\em U of Goettingen, Ibero-America Institute for Economic Research
  Discussion Paper}, (87), 2002.

\bibitem{Boss2008}
Michael~J Boss{\'e} and Johna Faulconer.
\newblock Learning and assessing mathematics through reading and writing.
\newblock {\em School Science and Mathematics}, 108(1):8--19, 2008.

\bibitem{Borasi1990}
Raffaella Borasi and Marjorie Siegel.
\newblock Reading to learn mathematics: New connections, new questions, new
  challenges.
\newblock {\em For the learning of mathematics}, 10(3):9--16, 1990.

\bibitem{Shaftel2006}
Julia Shaftel, Evelyn Belton-Kocher, Douglas Glasnapp, and John Poggio.
\newblock The impact of language characteristics in mathematics test items on
  the performance of english language learners and students with disabilities.
\newblock {\em Educational Assessment}, 11(2):105--126, 2006.

\bibitem{Freitag1997}
Mark Freitag.
\newblock Reading and writing in the mathematics classroom.
\newblock {\em The Mathematics Educator}, 8(1), 1997.

\bibitem{Bohlmann2002}
CA~Bohlmann and EJ~Pretorius.
\newblock Reading skills and mathematics: the practice of higher education.
\newblock {\em South African Journal of Higher Education}, 16(3):196--206,
  2002.

\bibitem{DiGisi1992}
Lori~Lyman DiGisi and Larry~D Yore.
\newblock Reading comprehension and metacognition in science: Status, potential
  and future direction.
\newblock 1992.

\bibitem{Anvari2002}
Sima~H Anvari, Laurel~J Trainor, Jennifer Woodside, and Betty~Ann Levy.
\newblock Relations among musical skills, phonological processing, and early
  reading ability in preschool children.
\newblock {\em Journal of experimental child psychology}, 83(2):111--130, 2002.

\bibitem{Hansen2002}
Dee Hansen and Elaine Bernstore.
\newblock Linking music learning to reading instruction.
\newblock {\em Music Educators Journal}, 88(5):17--52, 2002.

\bibitem{Zinar1976}
Ruth Zinar.
\newblock Reading language and reading music: Is there a connection?
\newblock {\em Music Educators Journal}, 62(7):70--74, 1976.

\bibitem{Brown2007}
Giorgina Brown, John Micklewright, Sylke~V Schnepf, and Robert Waldmann.
\newblock International surveys of educational achievement: how robust are the
  findings?
\newblock {\em Journal of the Royal statistical society: series A (statistics
  in society)}, 170(3):623--646, 2007.

\bibitem{Kolen2014}
Michael~J Kolen and Robert~L Brennan.
\newblock {\em Test equating, scaling, and linking: Methods and practices}.
\newblock Springer Science \& Business Media, 2014.

\bibitem{Liu2007}
Jinghua Liu and Michael~E Walker.
\newblock Score linking issues related to test content changes.
\newblock In {\em Linking and aligning scores and scales}, pages 109--134.
  Springer, 2007.

\bibitem{Stanley2018}
Christopher~T Stanley, Yaacov Petscher, and Hugh Catts.
\newblock A longitudinal investigation of direct and indirect links between
  reading skills in kindergarten and reading comprehension in tenth grade.
\newblock {\em Reading and Writing}, 31(1):133--153, 2018.

\bibitem{Bodovski2011}
Katerina Bodovski and Min-Jong Youn.
\newblock The long term effects of early acquired skills and behaviors on young
  children’s achievement in literacy and mathematics.
\newblock {\em Journal of Early Childhood Research}, 9(1):4--19, 2011.

\bibitem{Sparks2014}
Richard~L Sparks, Jon Patton, and Amy Murdoch.
\newblock Early reading success and its relationship to reading achievement and
  reading volume: Replication of ‘10 years later’.
\newblock {\em Reading and Writing}, 27(1):189--211, 2014.

\bibitem{Ding2009}
Cody~S Ding.
\newblock Measurement issues in designing and implementing longitudinal
  evaluation studies.
\newblock {\em Educational Assessment, Evaluation and Accountability},
  21(2):155--171, 2009.

\bibitem{Sakoe1978}
Hiroaki Sakoe and Seibi Chiba.
\newblock Dynamic programming algorithm optimization for spoken word
  recognition.
\newblock {\em IEEE transactions on acoustics, speech, and signal processing},
  26(1):43--49, 1978.

\bibitem{Vlachos2002}
Michail Vlachos, George Kollios, and Dimitrios Gunopulos.
\newblock Discovering similar multidimensional trajectories.
\newblock In {\em Proceedings 18th international conference on data
  engineering}, pages 673--684. IEEE, 2002.

\bibitem{Latecki2005}
Longin~Jan Latecki, Vasilis Megalooikonomou, Qiang Wang, Rolf Lakaemper,
  Chotirat~Ann Ratanamahatana, and Eamonn Keogh.
\newblock Elastic partial matching of time series.
\newblock In {\em European Conference on Principles of Data Mining and
  Knowledge Discovery}, pages 577--584. Springer, 2005.

\bibitem{Han2011}
Jiawei Han, Micheline Kamber, and Jian Pei.
\newblock {\em Data mining: Concepts and techniques}.
\newblock Morgan Kaufmann, 2011.

\bibitem{Glonek1995}
Garique~FV Glonek and Peter McCullagh.
\newblock Multivariate logistic models.
\newblock {\em Journal of the Royal Statistical Society: Series B
  (Methodological)}, 57(3):533--546, 1995.

\bibitem{Chiu2003}
Bill Chiu, Eamonn Keogh, and Stefano Lonardi.
\newblock Probabilistic discovery of time series motifs.
\newblock In {\em Proceedings of the ninth ACM SIGKDD international conference
  on Knowledge discovery and data mining}, pages 493--498, 2003.

\bibitem{Chan2005}
Philip~K Chan and Matthew~V Mahoney.
\newblock Modeling multiple time series for anomaly detection.
\newblock In {\em Fifth IEEE International Conference on Data Mining
  (ICDM'05)}, pages 8--pp. IEEE, 2005.

\bibitem{Wei2005}
Li~Wei, Nitin Kumar, Venkata~Nishanth Lolla, Eamonn~J Keogh, Stefano Lonardi,
  and Chotirat~(Ann) Ratanamahatana.
\newblock Assumption-free anomaly detection in time series.
\newblock In {\em SSDBM}, volume~5, pages 237--242, 2005.

\bibitem{Keogh2003}
E.~{Keogh}, J.~{Lin}, and W.~{Truppel}.
\newblock Clustering of time series subsequences is meaningless: implications
  for previous and future research.
\newblock In {\em Third IEEE International Conference on Data Mining}, pages
  115--122, 2003.

\end{thebibliography}

\clearpage
\begin{figure}[htbp]
  \centering
  \includegraphics[width=10cm]{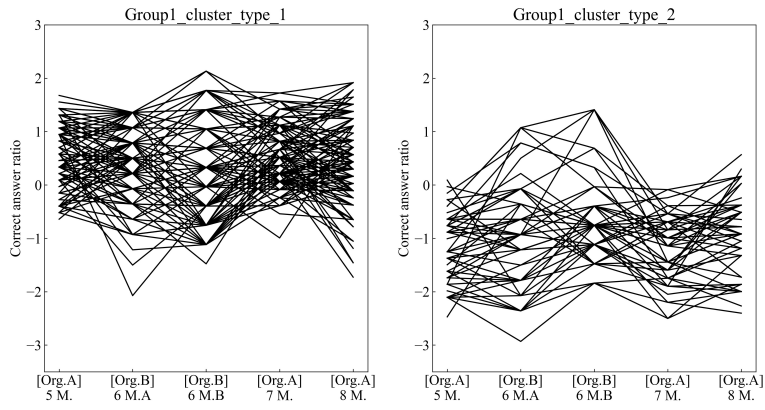}
  \caption{Group 1's results of Cluster Number 2}
  \label{fig:fig6}
\end{figure}
\begin{figure}[htbp]
  \centering
  \includegraphics[width=10cm]{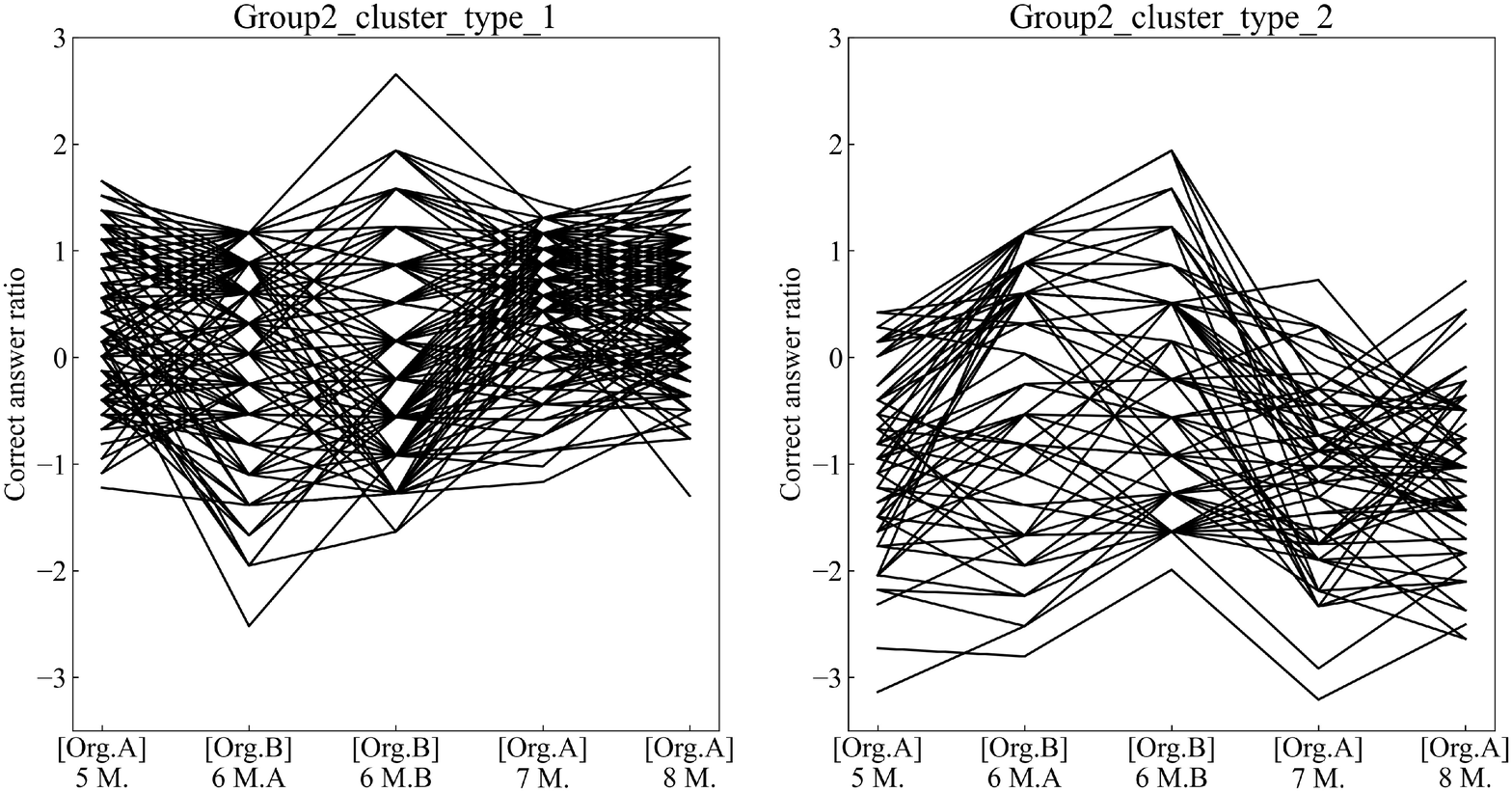}
  \caption{Group 2's results of Cluster Number 2}
  \label{fig:fig7}
\end{figure}
\begin{figure}[htbp]
  \centering
  \includegraphics[width=10cm]{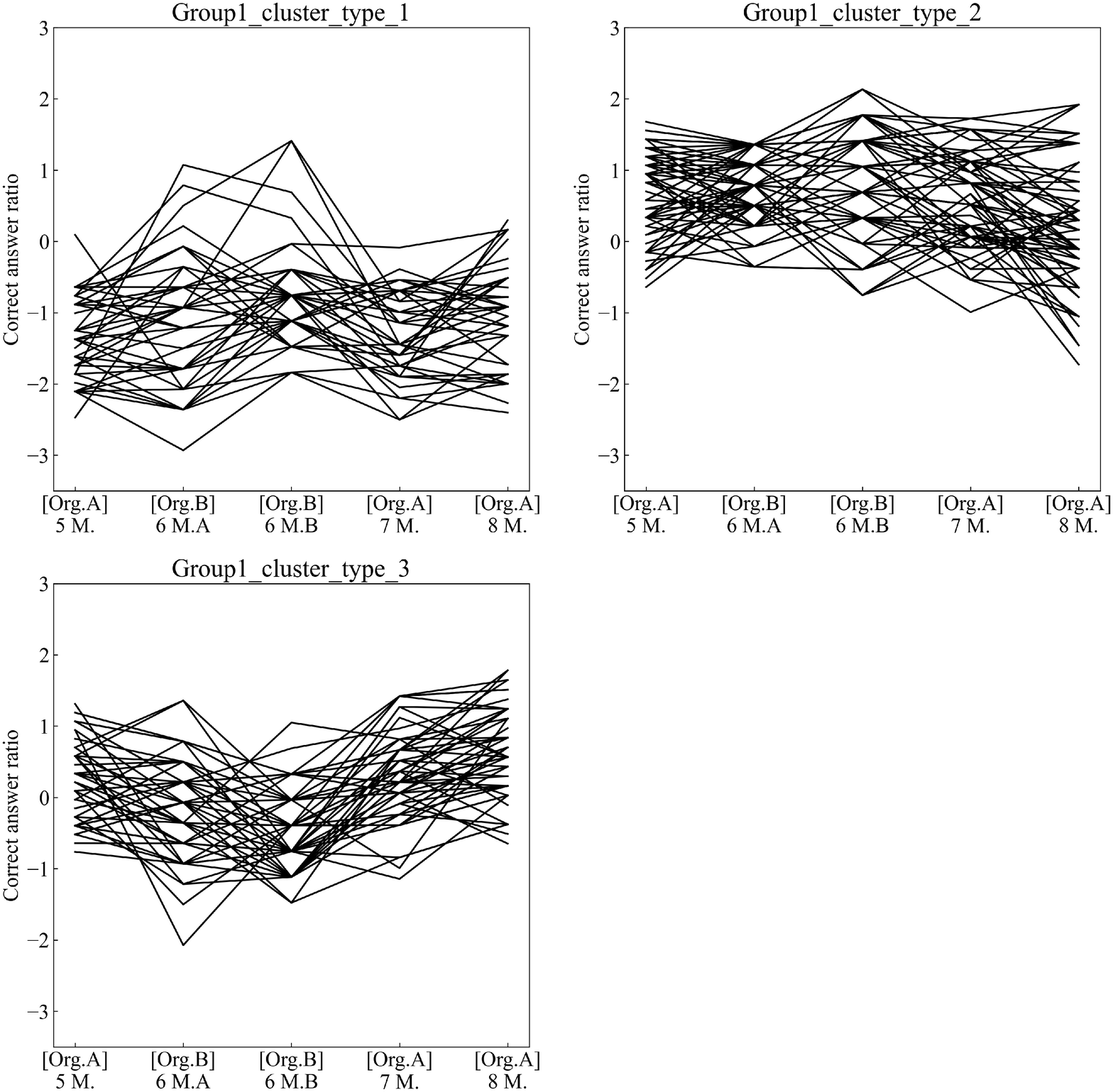}
  \caption{Group 1's results of Cluster Number 3}
  \label{fig:fig8}
\end{figure}
\begin{figure}[htbp]
  \centering
  \includegraphics[width=10cm]{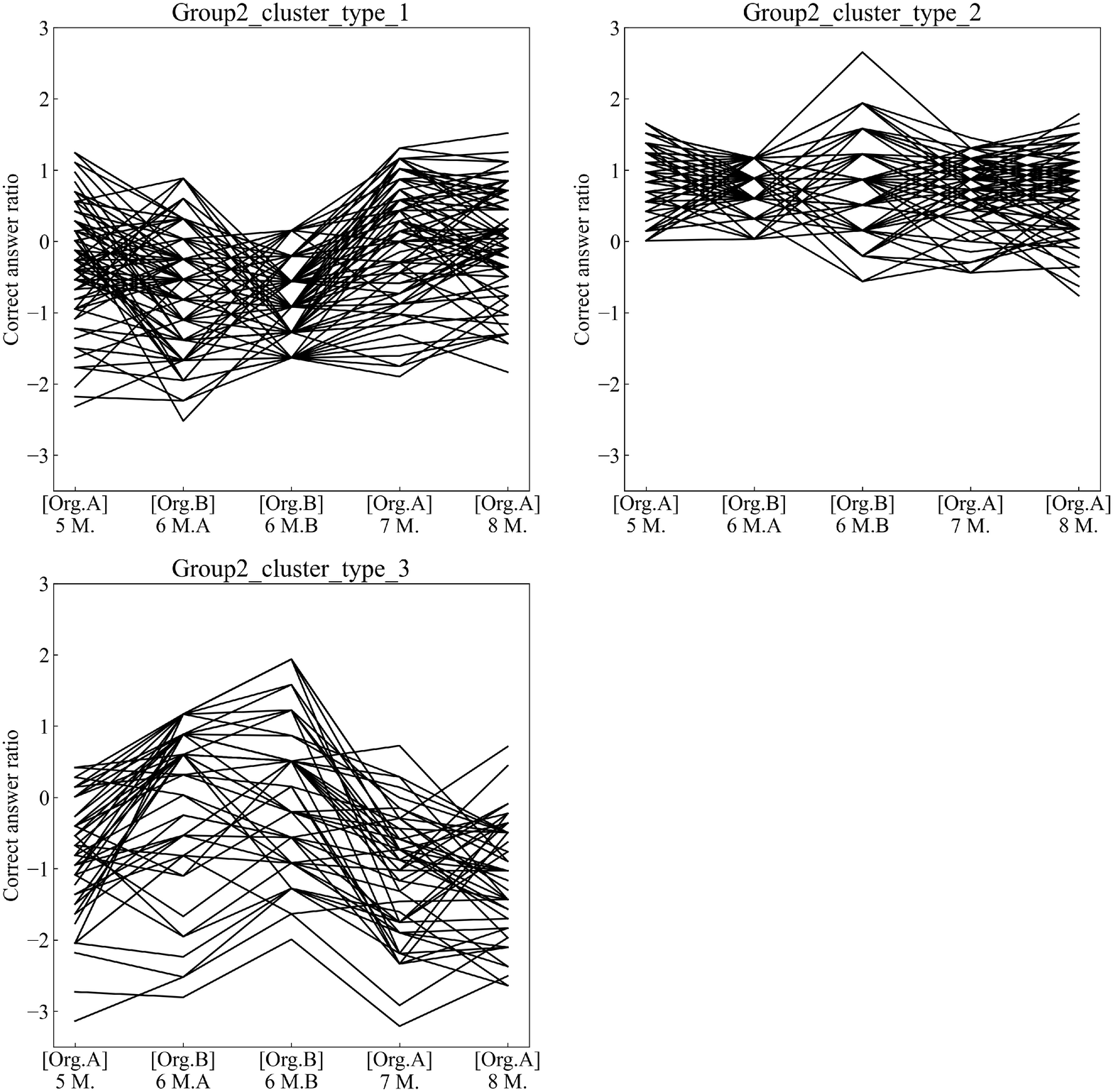}
  \caption{Group 2's results of Cluster Number 3}
  \label{fig:fig9}
\end{figure}
\begin{figure}[htbp]
  \centering
  \includegraphics[width=10cm]{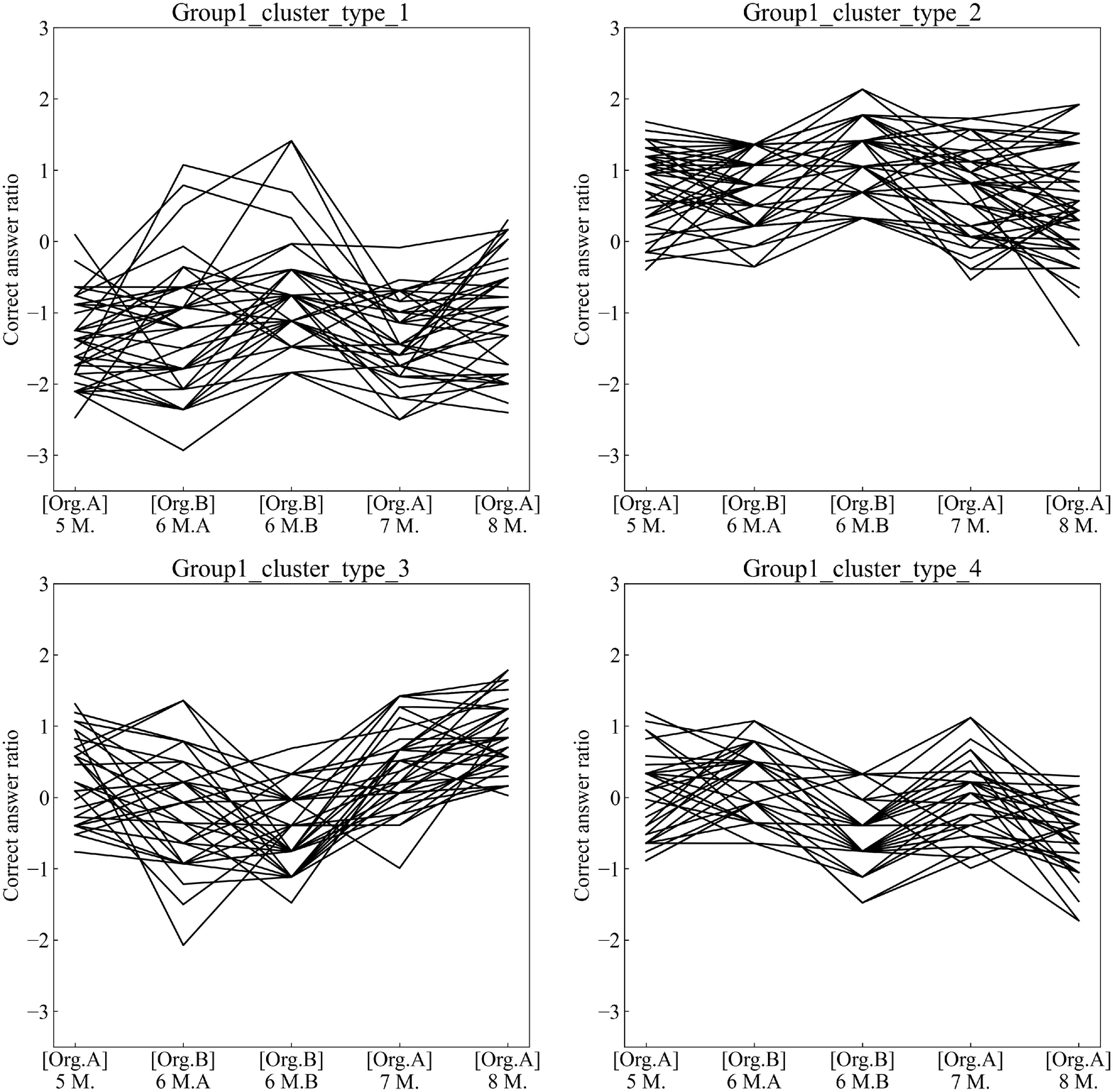}
  \caption{Group 1's results of Cluster Number 4}
  \label{fig:fig10}
\end{figure}
\begin{figure}[htbp]
  \centering
  \includegraphics[width=10cm]{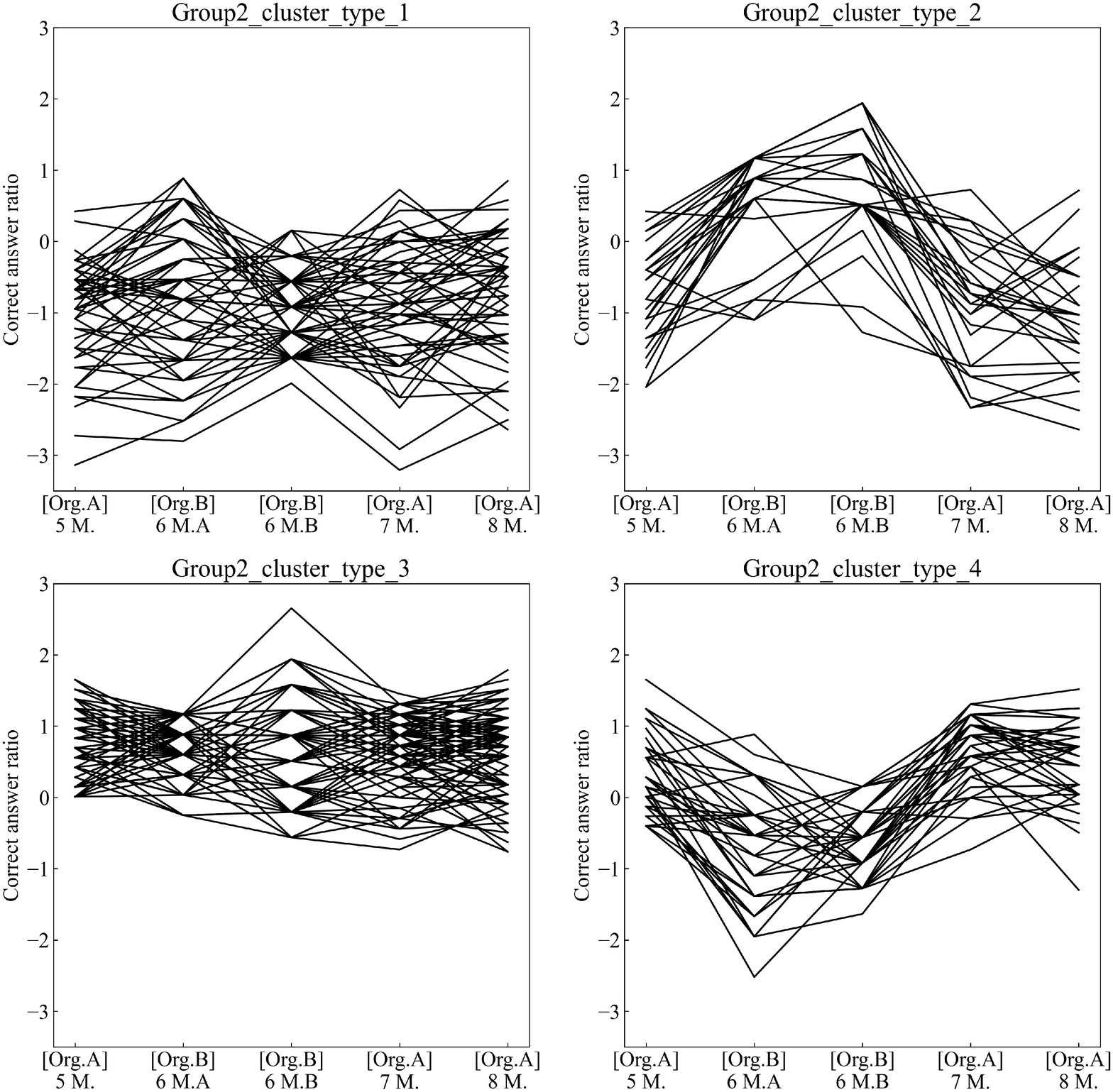}
  \caption{Group 2's results of Cluster Number 4}
  \label{fig:fig11}
\end{figure}
\begin{figure}[htbp]
  \centering
  \includegraphics[width=10cm]{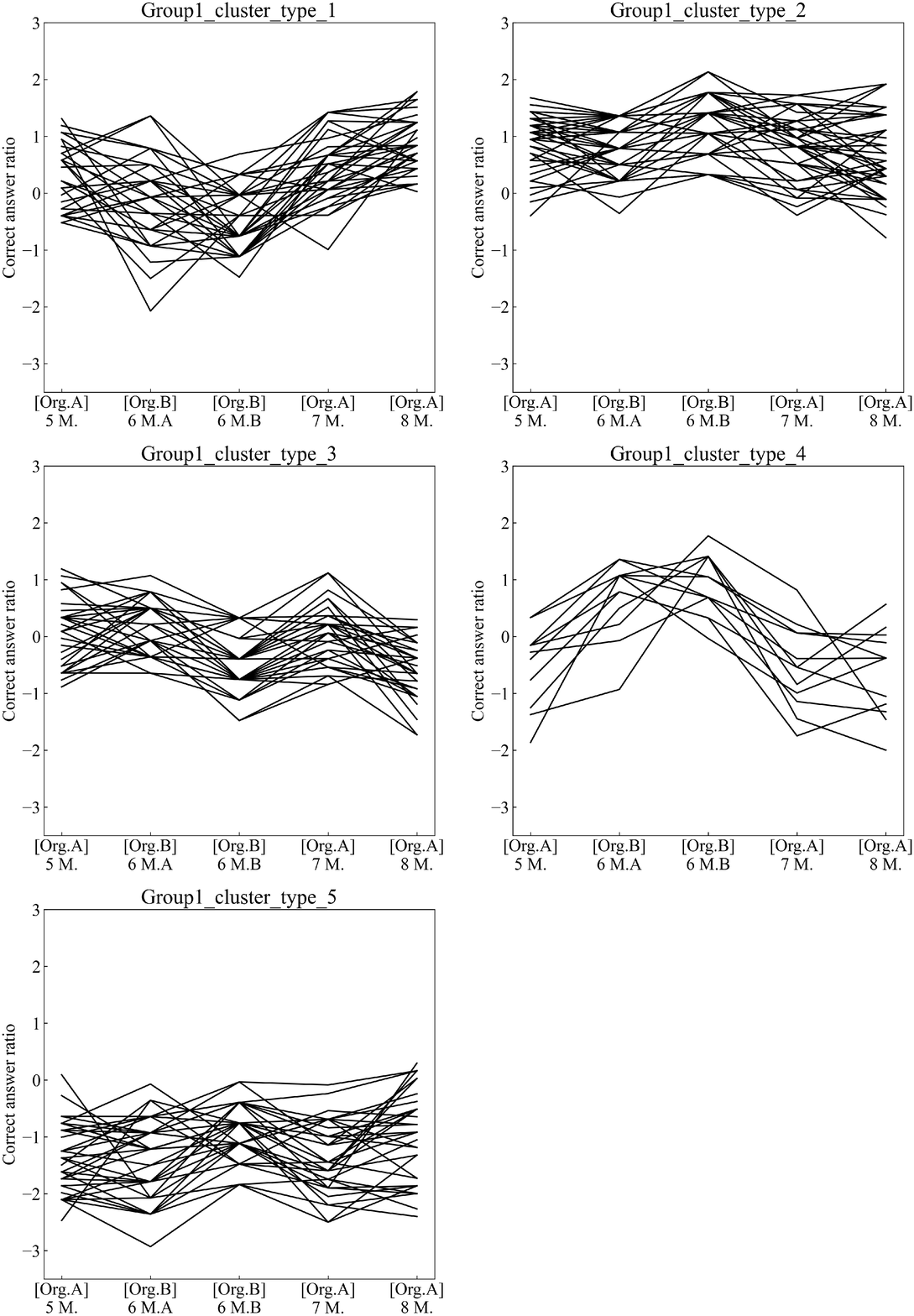}
  \caption{Group 1's results of Cluster Number 5}
  \label{fig:fig12}
\end{figure}
\begin{figure}[htbp]
  \centering
  \includegraphics[width=10cm]{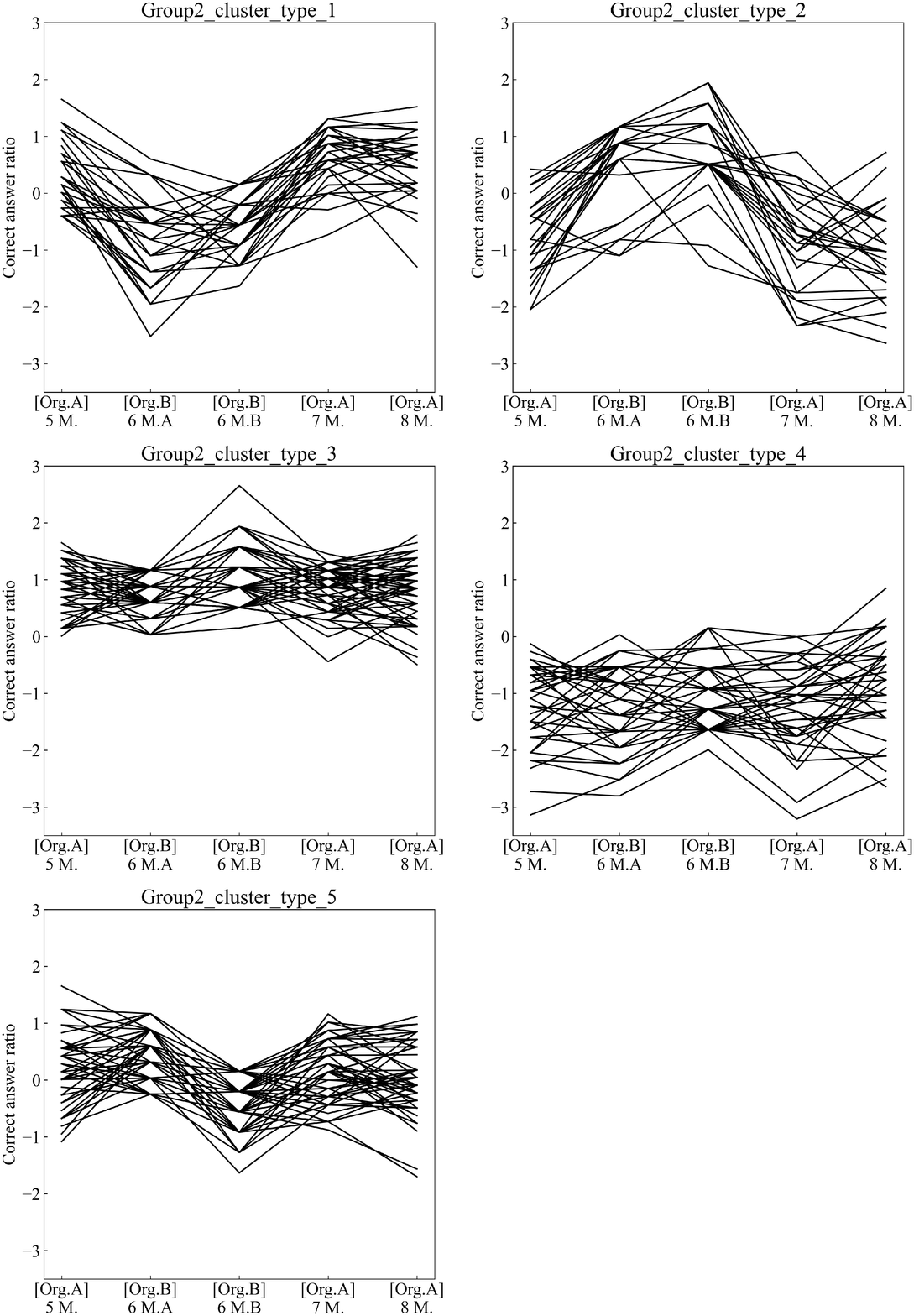}
  \caption{Group 2's results of Cluster Number 5}
  \label{fig:fig13}
\end{figure}
\begin{figure}[htbp]
  \centering
  \includegraphics[width=10cm]{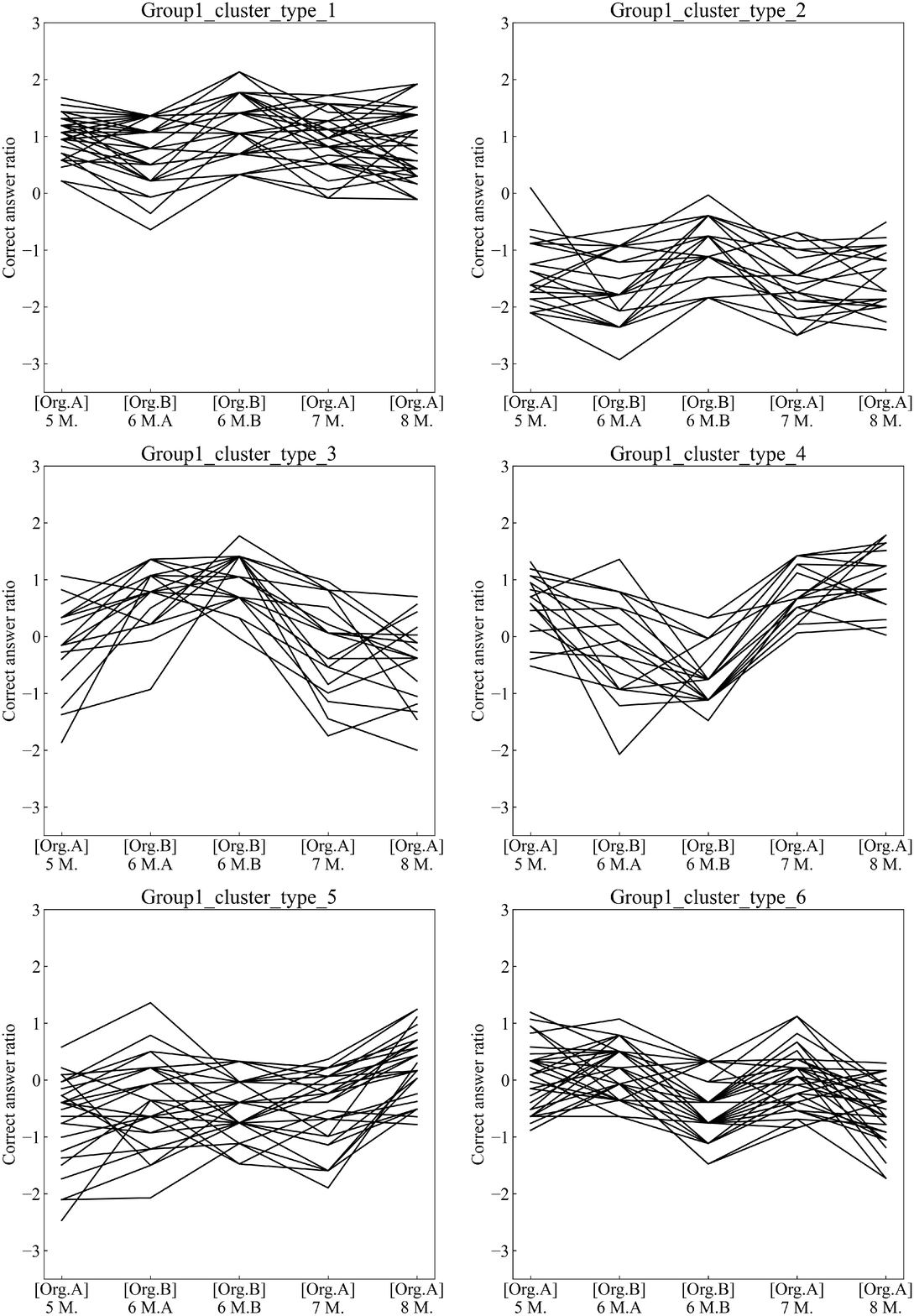}
  \caption{Group 1's results of Cluster Number 6}
  \label{fig:fig14}
\end{figure}
\begin{figure}[htbp]
  \centering
  \includegraphics[width=10cm]{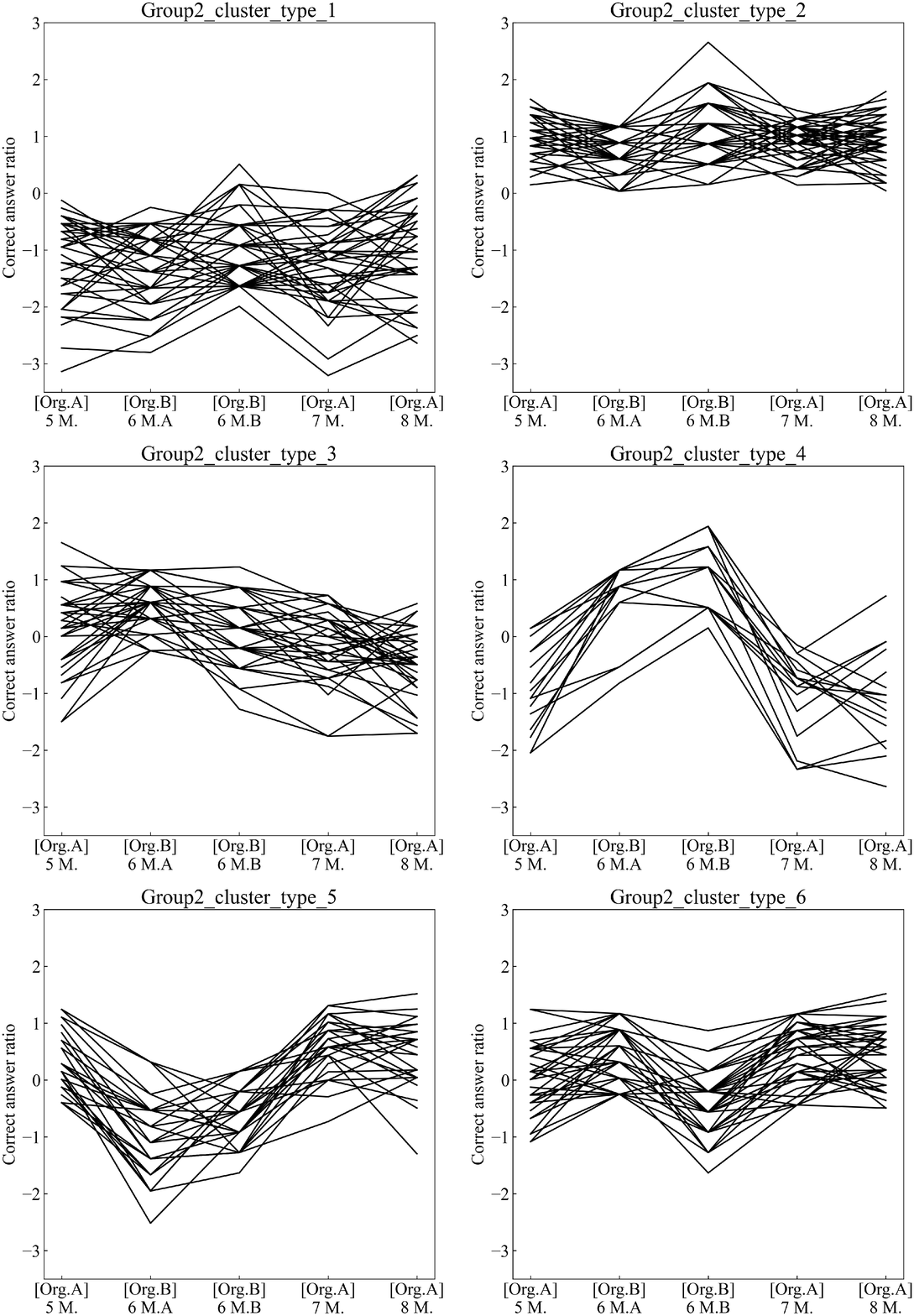}
  \caption{Group 2's results of Cluster Number 6}
  \label{fig:fig15}
\end{figure}
\end{document}